\documentclass[12pt,a4paper]{article}

\RequirePackage[l2tabu, orthodox]{nag}

\usepackage{mjheppub}
\usepackage{amsthm,amssymb,amsmath,epic,float}
\usepackage{rotating,epsfig,indentfirst,array,varioref}
\usepackage{appendix,tikz,mathtools}
\usepackage{setspace}
\usepackage{tikz}
\usetikzlibrary{decorations.pathreplacing}
\usetikzlibrary{calc}
\usetikzlibrary{positioning}
\usepackage{graphicx}

\let\emptyset\varnothing

\def\ll{\left\lgroup}
\def\rr{\right\rgroup}

\def\leq{\leqslant}
\def\geq{\geqslant}

\newcommand{\slthree}{\mathfrak{sl}_3}
\newcommand\mydef{\stackrel{\mathclap{\normalfont\mbox{def}}}{=}}

\newdimen\tableauside\tableauside=1.0ex
\newdimen\tableaurule\tableaurule=0.4pt
\newdimen\tableaustep
\def\phantomhrule#1{\hbox{\vbox to0pt{\hrule height\tableaurule
width#1\vss}}}
\def\phantomvrule#1{\vbox{\hbox to0pt{\vrule width\tableaurule
height#1\hss}}}
\def\sqr{\vbox{%
  \phantomhrule\tableaustep

\hbox{\phantomvrule\tableaustep\kern\tableaustep\phantomvrule\tableaustep}%
  \hbox{\vbox{\phantomhrule\tableauside}\kern-\tableaurule}}}
\def\squares#1{\hbox{\count0=#1\noindent\loop\sqr
  \advance\count0 by-1 \ifnum\count0>0\repeat}}
\def\tableau#1{\vcenter{\offinterlineskip
  \tableaustep=\tableauside\advance\tableaustep by-\tableaurule
  \kern\normallineskip\hbox
    {\kern\normallineskip\vbox
      {\gettableau#1 0 }%
     \kern\normallineskip\kern\tableaurule}%
  \kern\normallineskip\kern\tableaurule}}
\def\gettableau#1 {\ifnum#1=0\let\next=\null\else
  \squares{#1}\let\next=\gettableau\fi\next}

\def\be{\begin{equation}}
\def\ee{\end{equation}}
\def\ba{\begin{array}}
\def\ea{\end{array}}

\restylefloat{figure}

\newcommand{\cB}{\mathcal{B}}

\newcommand{\cG}{\mathcal{G}}

\newcommand{\cL}{\mathcal{L}}
\newcommand{\cM}{\mathcal{M}}

\newcommand{\cT}{\mathcal{T}}

\newcommand{\cW}{\mathcal{W}}

\def\ll{ \left\lgroup}
\def\rr{\right\rgroup}

\begin{document}

\title{\boldmath 
Rigid Fuchsian systems in  2-dimensional conformal field theories 
} 

\author[a\,\dagger]{Vladimir Belavin,}
\note[$\dagger$]{Weston Visiting Professorship at Weizmann Institute. On leave from
 Lebedev Physical Institute, Moscow.}
\author[b]{Yoshishige Haraoka,}
\author[c]{Raoul Santachiara\,}

\affiliation[a]{\vspace{2mm} 
Department of Particle Physics and Astrophysics, Weizmann Institute of Science,\\
Rehovot 7610001, Israel}
\affiliation[b]{\vspace{2mm} Department of Mathematics, Kumamoto University, Kumamoto, 860-8555, Japan}
\affiliation[c]{\vspace{2mm} LPTMS, CNRS (UMR 8626), Universit\'e Paris-Saclay, 15 rue Georges Cl\'emenceau,\\ 91405 Orsay,  France\vspace{2mm}}

\emailAdd{belavin@lpi.ru, haraoka@kumamoto-u.ac.jp, raoul.santachiara@u-psud.fr
}


\abstract{
We investigate  Fuchsian equations arising in the context of 2-dimensional conformal field theory (CFT) and we apply the Katz theory of Fucshian {\it rigid} systems to solve some of these equations. We show that the Katz theory provides a precise mathematical framework to answer the question whether the fusion rules of degenerate primary fields are enough for determining the differential equations satisfied by their correlation functions.
We focus on the case of $\cW_3$ Toda CFT: we argue that the differential equations arising for four-point conformal blocks with one $n$th level semi-degenerate field and a fully-degenerate one in the fundamental $sl_3$ representation are associated to Fuchsian  rigid systems. We show how to apply Katz theory to determine the explicit form of the differential equations, the integral expression of solutions and the monodromy group representation. The theory of twisted homology is also used in the analysis of the integral expression. This approach allows to construct the corresponding fusion matrices and to perform the whole bootstrap program: new explicit factorization of $W_3$ correlation functions as well shift relations between structure constants are also provided.
}

\maketitle
\flushbottom

\section{Introduction}
The symmetry algebra of any 2-dimensional conformal field theory (CFT) is the tensor product $\mathfrak{L}\otimes \bar{\mathfrak{L}}$ of two algebras generated by a set of conserved currents, which include as sub-algebras the holomorphic $\mathfrak{V}_c$ and anti-holomorphic $\bar{\mathfrak{V}}_c$ Virasoro algebras of central charge $c$, $\mathfrak{V}_c\subseteq  \mathfrak{L}$ and  $\bar{\mathfrak{V}}_c\subseteq \bar{\mathfrak{L}}$~\cite{zam85}. Primary fields  $\Phi_{\alpha,\alpha'}(z,\bar{z})$, associated to highest weight vectors in the $\mathfrak{L}\otimes \bar{\mathfrak{L}}$ representations,  are generally labeled by a pair of vector parameters $ \alpha$ and $\alpha'$. Other local (descendant) fields are obtained by acting with the  modes of chiral and anti-chiral currents on the primary fields. 
The $n$-point correlation functions $\left< \prod_{i=1}^n \Phi_{\alpha_i,\alpha'_i}(z_i,\bar{z}_i)\right>$ are built of the so-called conformal blocks. The conformal blocks arise as special functions in the representation theory of the $\mathfrak{A}$ and $\bar{\mathfrak{A}}$ algebras. The simplest correlation function that encodes all information about a particular CFT model is the four-point function, whose decomposition takes the form
\begin{equation} 
\left< \prod_{i=1}^4 \Phi_{\alpha_i,\alpha'_i}(z_i,\bar{z}_i)\right>= \sum_{M,M'} H_{M,M'}\cB_{M}(\{z_i|\alpha_i\})\bar{\cB}_{M'}(\{\bar{z}_i|\alpha'_i\})\;.\label{corr_decomp}
\end{equation}
Using  global conformal invariance, the 4-point conformal blocks $\cB_{M}(\{z_i|\alpha_i\})$ ($\bar{\cB}_{M'}(\{\bar{z}_i|\alpha'_i\})$)  can be studied as functions of the cross ratios $z$ ($\bar{z}$), $\cB_{M}(\{z_i|\alpha_i\})\to \cB_{M}(z|\{\alpha_i\})$  and $\bar{\cB}_{M'}(\{\bar{z}_i|\alpha'_i\}) \to \bar{\cB}_{M'}(\bar{z}|\{\alpha'_i\})$. They depend on the parameters $\{(\alpha_i,\alpha'_i)\}$ of the fields entering the correlation function (the external fields) and on the internal parameters $M, M'$. The internal parameters $(M, M')$  specify the representation $(\alpha_M, \alpha'_M)$ produced in some chosen fusion channel, for instance when $z_1\to z_2$,  $\Phi_{\alpha_1,\alpha'_1}(z_1,\bar{z}_1)\Phi_{\alpha_2,\alpha'_2}(z_2,\bar{z}_2) \sim \Phi_{\alpha_M,\alpha'_M}(z_2,\bar{z}_2)$. 
In general, the knowledge  
of $\alpha_M$ ($\alpha'_M$)  is not sufficient to determine uniquely the conformal block   
$\cB_{M}(\{z_i|\alpha_i\})$ ($\bar{\cB}_{M'}(\{\bar{z}_i|\alpha'_i\})$) and  a possibly infinite set of additional parameters $\lambda_M$ ($\lambda'_M$) has to be specified.  Therefore,  the notation  $M$ and $M'$ stands for the whole set of the parameters that are necessary to uniquely define the conformal block, $(M, M')=(\{\alpha_M,\lambda_M\}, \{\alpha'_M,\lambda'_M\})$. The constants $H_{M,M'}$,  can be directly related to the structure constants of the theory and are determined by demanding some global properties of the correlation functions. For instance, a very natural condition is to impose that the correlation function is single-valued in the complex plane, i.e. it has a simple monodromy. 
This is the basis of the conformal bootstrap approach.  At the best of our knowledge, apart from a particular case considered in \cite{BEFS16}, the explicit decomposition  (\ref{corr_decomp}) is not known in the presence of additional  parameters $\lambda_M$ ($\lambda'_M$). In this paper we will give new insights concerning this problem by using the theory of rigid Fuchsian systems \cite{katz1996rigid}.  

In the case where the chiral algebra coincides with the Virasoro one, $\mathfrak{L}=\mathfrak{V}_c$,\footnote{In what follows we concentrate on the holomorphic sector,  since the anti-holomorphic one  can be considered similarly.} the representations are uniquely characterized by their conformal dimensions $h$. 
In terms of the above notations then  $\alpha =h$.  The $z$-series expansion of $\cB_M (z, \{h_i\})$ is determined by the matrix elements between primary fields (specified more in detail below) which in turn are completely determined as functions of the $4$ external  conformal dimensions $h_{i}$, and on the dimension $h_{M}$ of the internal channel. 
Therefore there are no additional  parameters and one can set $M =h_M$. Even though there are various efficient methods to compute the Virasoro conformal block's series expansion (see, e.g. \cite{zam87b}) and the analytic continuation of this series is also known \cite{pt00}, 
the Virasoro conformal block remains a complicated object and its closed form is not known in general. In this respect a crucial observation \cite{bpz84} was that $\cB_{h_M} (z,\{h_i\})$ satisfies an ordinary differential equation (ODE) of Fuchsian type if at least one value $h_i$ corresponds to a degenerate representation. A representation  with dimension $h$ is said to be degenerate at level $L$ if it contains a null-state of dimension $h+L$. For instance, if $h_1=0$, the corresponding identity field is degenerate at level one. This implies that $\partial_z \cB_{h_M} (z,\{0,h_i\})=0$, which simply constraints the conformal block to have no dependence on the position of the identity field. The first non-trivial  ODE has been  derived from the degenerate representation at level two. This representation plays a crucial role,  since the fusion of level two degenerate fields produces all possible higher level degenerate fields. Accordingly, the corresponding second-order ODE is of fundamental importance and has numerous applications at the heart of the development of the Virasoro theory.  We mention for instance the construction of  the solution  of the Liouville field theory \cite{te95}, its recent  mathematical rigorous derivation \cite{vargasdozz17}, the SLE-CFT \cite{Bauer2003} connection, as well as the integrable structure of CFTs \cite{epss11}.  This second-order ODE takes the form of a Gauss hypergeometric equation and represents a simplest {\it rigid Fuchsian system}. The rigid Fuchsian systems are introduced below and their application to CFTs algebras will be the focus of this paper.

One of the natural generalizations of the Virasoro algebra is the $\mathcal{W}_3$ algebra~\cite{zam85,fl88,bs95}, $\mathfrak{L}= \mathcal{W}_3$, that have found  important applications in the fractional quantum Hall physics \cite{ReRe99,es09,epss11} and critical statistical models \cite{fl88,DEI16}. In addition to the holomorphic stress-energy tensor $\cT(z)$, this algebra is generated by a current $\cW(z)$ of spin 3. The representation of the  $\mathcal{W}_3$ algebra is characterized by a pair of parameters $(h, q)$  respectively representing the conformal dimension and the $\cW(z)$ zero-mode eigenvalue of the primary field (using the previous notation $\alpha= (h, q)$).  
The $\mathcal{W}_3$ theory is well known to be an upgrade in difficulty level compared to Virasoro case. Here the general conformal block is not determined by  specifying the representations of the external fields  $(h_i,q_i)$ and of the internal channel $(h_M,q_M)$ because $\mathcal{W}_3$ symmetry constraints are not strong enough  to fix all matrix elements, which contribute to the expansion coefficients of the conformal block. In general $\mathcal{W}_3$ conformal block depends on an infinite set of additional parameters $\lambda_1,\lambda_2,\cdots$ and one should think of the internal parameter $M$  as containing  this set, $M= h_M,q_M, \lambda_1,\lambda_2,\cdots$. This is the reason why, differently from the Virasoro case, the series expansion of $\cW_3$  conformal blocks can be given only for a particular choice of external representations. In \cite{fl07}  the conformal block containing a fully degenerate field \footnote{Precise definitions concerning different types of  $\cW_3$ representations are given in sec.~\ref{section.02.w3.toda.conformal.field.theory}.} associated to the fundamental  $\slthree$ representation  {\it and} a semi-degenerate field at level one has been studied. In this case the conformal block has no additional parameters, admits a combinatorial representation  in terms of Nekrasov functions \cite{wyl09,mironov2010agt,Alkalaev:2014sma, bfs15} and satisfies recurrence relations \cite{Pogrec17}. This conformal block satisfies a third-order Fuchsian equation \cite{fl07c,rib08b} and has been used to compute all the matrix elements and structure constants involving a semi-degenerate field at level one. Analogously to the case of the second-order ODE appearing in the Virasoro algebra, this third-order ODE has a fundamental nature and as discussed below is also related to a rigid Fuchsian system. More recent results on $\cW_3$ (and more in general on $\cW_N$) conformal blocks have been obtained in \cite{Gav15,BeGaMa17} for integer values of the central charge.   

In a recent study some of the authors \cite{BCES17} considered the case in which the conformal block contains a semi-degenerate field at level 2. This case is interesting because here we meet conceptually new situation -- appearance of the multiplicities in the fusion channels. As a consequence, this conformal block contains one additional parameter $\lambda_1$, $M=(h_M, q_M, \lambda_1)$. It was shown that it satisfies a sixth-order Fuchsian ODE, however the solutions of this equation was not given. Hence, the monodromy group has not been  available. We note that the knowledge of the monodromy group allows to construct the correlation function (\ref{corr_decomp}) with the required global properties. In this paper we provide the solution to this sixth-order ODE. 

The crucial observation is that the sixth-order ODE is also related to a Fuchsian rigid system. We show how to apply the  Katz theory to find the explicit form of the differential equations, the integral expression of the solutions and the monodromy group. In particular we provide the explicit decomposition (\ref{corr_decomp}) and prove its uniqueness.
Finally, we consider a more general situation where the conformal block contains one fundamental field and one semi-degenerate field at level $n$. In this case there are $n-1$ additional parameters. We argue that this conformal block satisfies a $3n$-th order ODE, which is also related to a  rigid system. This series of $3n$-th order differential ODEs can be considered as  the fundamental set of ODEs in the $\cW_3$ theory, on the same footing as the Gauss-hypergeometric equation in the Virasoro theory.

The paper is organized as follows. In section~\ref{section.02.w3.toda.conformal.field.theory} we introduce basic notations in  $\cW_3$ theory and we define the  $\cW_3$ conformal blocks. We focus our attention on the conformal blocks that contain one fully-degenerate fundamental field and a semi-degenerate field at level $n$. We derive the order and the local exponents  of the differential equations satisfied by these family of functions: these equations are associated to Katz rigid systems. In section ~\ref{diffeq} we discuss a set of rigid Fuchsian systems that appear in CFT and review their solutions in the solved case. In section~\ref{Katz-theory} we briefly review the Katz theory on rigid local systems, and then show how to apply it to reproduce known results in CFT. In section ~\ref{applications2}  we apply the Katz theory to solve the $6$-th order ODE found in \cite{BCES17} and we provide new explicit factorizations (\ref{corr_decomp}) for correlation function with additional parameters.
In section~\ref{summary} we resume the results and discuss some future perspectives.

\section{Differential equations for $\cW_3$ conformal blocks}

\label{section.02.w3.toda.conformal.field.theory}

In this section we recall basic elements of  $\cW_3$ CFT necessary for our forthcoming discussion (for reviews, see, e.g., \cite{fl90} and \cite{bs92}). 
For more detailed discussion of the  $\cW_3$ chiral algebra, its representation modules and its conformal blocks, consistent with the notations and normalizations used here, the reader is referred  to \cite{BEFS16,BCES17}. 

\subsection{Representations of $\cW_3$ algebra and primary fields}


The  $\cW_3$ algebra is the associative algebra generated by the spin-$2$ stress-energy tensor $\cT (x)$ and the spin-$3$ holomorphic field $\cW(x)$.

A convenient parametrization of the $\cW_3$ algebra central charge is
\begin{equation}
c = 2 + 24 \;Q^2\;, \quad 
Q = b + \frac{1}{b}\;.
\end{equation}
A highest weight representation $\cM_{h,q}$ of the $\cW_3$ algebra is labeled by the parameters $h$ and $q$, representing respectively $\cT(x)$ and $\cW(x)$ zero modes eigenvalues. Alternatively, $\cM_{h,q}$ representation can be characterized by a bi-dimensional vector 
$\vec{\alpha}$ defined in the basis of the  fundamental $\slthree$ weights $\vec{\omega_i}$ with $i=1,2$:
\begin{equation}
\vec{\alpha} = \alpha_1 \;\vec{\omega}_1+\alpha_2 \;\vec{\omega}_2\;.
\end{equation}
Here and below we use the following $\slthree$ data
\begin{align}
\vec{\omega}_1 ={}&  \sqrt{\frac23} \left(1, 0\right)\;,\quad  \vec{\omega}_2 = \sqrt{\frac16} \left(1, \sqrt{3}\right)\;,\\
\vec{e}_1 = {}& 2 \vec{\omega}_1 - \vec{\omega}_2\;,\quad 
\vec{e}_2 =  -\vec{\omega}_1 + 2 \vec{\omega}_2\;,
\quad \vec{\rho}  = \vec{\omega}_1 + \vec{\omega}_2\;,\\
\vec{h}_1 ={}&  \vec{\omega}_1\;,\quad 
\vec{h}_2 =  \vec{\omega}_1 - \vec{e}_1\;,\quad
\vec{h}_3 =  \vec{\omega}_1 - \vec{e}_1 - \vec{e}_2\;,
\end{align}
where $e_{k}$ are the simple roots, $\rho$ is the Weyl vector (half the sum of all positive roots) and $h_{i}$ are the weights of the fundamental representation.

The relation between $(h,q)$ and $\vec{\alpha}$ parametrization is given by
\begin{equation}
	h  = Q^2 + x_1 x_2 + x_1 x_3 + x_2 x_3\quad \text{and} \quad q = i \; x_1 x_2 x_3
\end{equation}
with
\begin{equation}
x_i =  \left(Q\vec{\rho}-\vec{\alpha}\right) \cdot \vec{h}_i\;. 
\end{equation}
\noindent The above expressions are invariant under the $\slthree$ Weyl group action. 
This group is composed by six elements $\hat{s}(\vec{\alpha})$ which act on $\vec{\alpha}=(\alpha_1,\alpha_2)$ 
in the following way:
\begin{eqnarray}
 (\alpha_1,\alpha_2) &\mapsto & 
 (\alpha_1,\alpha_2),\
 (2 Q-\alpha_1, \alpha_1+\alpha_2-Q),\ 
 (\alpha_2, -\alpha_1 - \alpha_2+Q),\nonumber \\ \label{weyl}&&
 (2Q-\alpha_2, 2 Q-\alpha_1),\ 
 (\alpha_1 + \alpha_2-Q, 2 Q-\alpha_2),\  (-\alpha_1 - \alpha_2+Q, \alpha_1).
\end{eqnarray} 
Henceforth we use two equivalent ways of labeling primary fields, corresponding to the highest weight  vectors of  $\cW_3$ modules: $\Phi_{h,q}(x)$ or $\Phi_{\vec{\alpha}}(x)$.

The descendant fields $\Phi^{(I)}_{\vec{\alpha}}$ are obtained by acting with the Laurent modes $L_{-n}$ and $W_{-m}$ of $\cT(x)$ and $\cW(x)$ on the primary fields: 

\begin{equation}
\Phi^{(I)}_{\vec{\alpha}}\mydef \cL_I\;\Phi_{\vec{\alpha}} \mydef\; L_{-i_m} \cdots L_{-i_1} 
                            W_{-j_n} \cdots W_{-j_1} \Phi_{\vec{\alpha}}\;.
\label{W3Verma}                            
\end{equation}
Here $I$ stands for the sets of ordered positive integers $
I=\{i_m, \cdots, i_1 ; j_n, \cdots, j_1\}$ with $i_m \geq \cdots \geq i_1 \geq 1, \quad 
j_n \geq \cdots \geq j_1 \geq 1$. The level (minus the sum of the mode numbers) defines a natural $L_0$ grading in the $\cM_{\vec{\alpha}}$ module, $|J|=\sum_r i_r +\sum_s j_s$. The symbol $\emptyset$ will be used when no modes $L_{i}$ or $W_j$ are present, $\Phi^{(\emptyset)}_{\vec{\alpha}}=\Phi_{\vec{\alpha}}$.

The conformal block $\cB_{M}\ll x|\vec{\alpha}_{1},\vec{\alpha}_{2},\vec{\alpha}_{R},\vec{\alpha}_{L}\rr$ is defined by the following expansion:
\begin{multline}
x^{h_L + h_R - h_M} \cB_M \ll x|L,2,1,R\rr
=
\\
1 + 
\sum_{l=1}^\infty x^l 
\sum_{
\substack{
K, K' 
\\ 
|K| = |K'| = l
}
} 
 \left[H^{(l)}\ll \vec{\alpha}_M\rr\right]^{-1}_{K, K'} \Gamma_{\emptyset, \emptyset, K}\ll L,2,\vec{\alpha}_M\rr
 \Gamma^{\prime}_{K', \emptyset, \emptyset}\ll \vec{\alpha}_M,1,R\rr,
\label{cb.series.expansion}
\end{multline}
\noindent where $L,R,2,1$ is a short notation for  $\vec{\alpha}_{L},\vec{\alpha}_R,  \vec{\alpha}_{2}, \vec{\alpha}_{1}$, $H^{(l)}_{K,K'}(\vec{\alpha})=\langle (\Phi^{*}_{\vec{\alpha}})^{K}|\Phi_{\vec{\alpha}}^{K'}\rangle$ is  the matrix of scalar products between the descendants at level $l$ and $\Gamma, \Gamma'$ are normalized matrix elements:
\begin{eqnarray}
\label{gamma-gammap} 
\Gamma_{I, J, K} \ll \vec{\alpha}_L,\vec{\alpha}_M,\vec{\alpha}_R\rr
 &=&
\frac{\langle 
(\Phi^{*}_{\vec{\alpha}_L})^{(I)}| 
\, 
\Phi^{(J)}_{\vec{\alpha}_M}(1) 
\, 
\Phi^{(K)}_{\vec{\alpha}_L}(0)
\rangle}{\langle 
(\Phi^{*}_{\vec{\alpha}_L})| 
\, 
\Phi_{\vec{\alpha}_M}(1) 
\, 
\Phi_{\vec{\alpha}_R}(0)
\rangle}\;,\\
\Gamma^{\prime}_{I, J, K} \ll \vec{\alpha}_L,\vec{\alpha}_M,\vec{\alpha}_R\rr &=& 
\frac{\langle 
\Phi^{(I)}_{\vec{\alpha}_L}| 
\, 
\Phi^{(J)}_{\vec{\alpha}_M}(1) 
\, 
\Phi^{(K)}_{\vec{\alpha}_R}(0)
\rangle}{\langle 
\Phi_{\vec{\alpha}_L}| 
\, 
\Phi_{\vec{\alpha}_M}(1) 
\, 
\Phi_{\vec{\alpha}_R}(0)
\rangle} \;.
\end{eqnarray}
In the above expressions we defined the conjugate field $\Phi^{*}_{\vec{\alpha}}\mydef \Phi_{2Q \vec{\rho}- \vec{\alpha}}$.\footnote{The conjugation takes into account the diagonal form of the 2-point function $\langle\Phi^{*}_{\vec{\alpha}}(1)\Phi_{\vec{\alpha}'}(0)\rangle=\delta_{\vec{\alpha},\vec{\alpha}'}$.}

It can be shown \cite{bs92,kms10} that any matrix element can be written as a linear 
combination of the following matrix elements
\begin{equation}
\lambda^{(l)}_{\vec{\alpha}_L,\vec{\alpha}_{R}}(\vec{\alpha}_M)\mydef \Gamma_{\emptyset, \{\emptyset;\underbrace{1,\cdots,1}_{l}\}, \emptyset} \ll \vec{\alpha}_L,\vec{\alpha}_M,\vec{\alpha}_R\rr =\frac{\langle \Phi^*_{\vec{\alpha}_L} (\infty)| (W_{-1}^l\Phi_{\vec{\alpha}_M})(1) \Phi_{\vec{\alpha}_R} (0)\rangle}{\langle \Phi^*_{\vec{\alpha}_L} (\infty)| \Phi_{\vec{\alpha}_M}(1) \Phi_{\vec{\alpha}_R} (0)\rangle}\;,
\label{basis_me}
\end{equation}
where $ l= 1, 2, \cdots$. For general $\vec{\alpha}_M$, the matrix element $\lambda^{(l)}(\vec{\alpha}_M)$ is unknown and remains to be a free parameter.
 
\paragraph{Fully and semi-degenerate representations and fusion rules.}
\label{degeneratefields} 
A  {\it fully-degenerate} $\cW_3$ representation is associated with the primary 
field $\Phi_{\vec{\alpha}_{r_1 r_2 s_1 s_2}}(x)$ with the parameter
\begin{equation}
\vec{\alpha}_{r_1 r_2 s_1 s_2} = 
b
\ll
(1 - r_1)\; \vec{\omega}_1 +
(1 - r_2)\; \vec{\omega}_2 
\rr 
+ 
\frac{1}{b}
\ll
(1 - s_1)\; \vec{\omega}_1 +
(1 - s_2)\; \vec{\omega}_2 
\rr,
\end{equation}
where  $r_1, r_2, s_1,s_2$  are positive integers. 
This representation exhibits two independent null-states at levels $r_1 s_1$ and  $r_2 s_2$ \cite{fl90} and the fusion products of the field $\Phi_{\vec{\alpha}_{r_1 r_2 s_1 s_2}}$ with a general primary field $\Phi_{\vec{\alpha}}$ takes the form
\begin{equation}
\label{w3: fuprod}
\Phi_{\vec{\alpha}_{r_1 r_2 s_1 s_2}} \times \Phi_{\vec{\alpha}} =  
\oplus_{\vec{h}_r, \vec{h}_s} \Phi_{\vec{\alpha} - b \vec{h}_r-b^{-1} \vec{h}_s}\;,
\end{equation}
where $h_r$ and $h_s$ are the weights of the $\slthree$ representations with the highest-weights: \\$(r_1-1)\; \vec{\omega}_1 +
( r_2-2)\; \vec{\omega}_2$ and $(s_1-1)\; \vec{\omega}_1 +
( s_2-2)\; \vec{\omega}_2$ respectively.

A primary field $\Phi_{\vec{\alpha}}=\Phi_{(1-n) b \vec{\omega}_1+ s \vec{\omega}_2}$, where $s$ is  an arbitrary complex number, is called {\it semi-degenerate}  at level $n$.  The corresponding representation contains a null-vector at level $n$ whose  explicit expression  has been found in \cite{bw92}. In this case, for general $\vec{\alpha}_{L,R}$, one has \cite{bw92,BCES17} the following set of free parameters:
\begin{equation}
\begin{aligned}
\lambda^{(i)}_{\vec{\alpha}_L,\vec{\alpha}_R}((1-n) b \vec{\omega}_1+ s \vec{\omega}_2), \quad i=1,\dots,n-1  &\quad\to& \quad \text{undetermined} 
\\
\lambda^{(i)}_{\vec{\alpha}_L,\vec{\alpha}_R}((1-n) b \vec{\omega}_1+ s \vec{\omega}_2), \quad i=n,n+1,\dots  &\quad\to& \quad \text{fixed}
\label{param}
\end{aligned}
\end{equation}
The latter are expressed  in terms of the former by using the vanishing of the null-vector.

\subsection{Conformal blocks with semi-degenerate fields}

\label{cb_sem_deg}

We consider the conformal block $\cB^{(n,s)}_{M} (x)\mydef \cB_{M} (x|-b\vec{\omega}_1,-n b\vec{\omega}_1+ s\vec{\omega}_2,\vec{\alpha}_L,\vec{\alpha}_R)$, that can be represented by the following diagram:
\begin{equation}
\begin{aligned}
\label{cbconsidered}
\begin{tikzpicture}
[line width=1.2pt]
\draw (-2.5,.5)node[left]{$\cB^{(n,s)}_M (x)$};
\draw (-2.5,.5) node[right]{$\mydef$};
\draw (0,0)--(1,0);
\draw (1,0)--(1,1);
\draw (1,0)--(3,0);
\draw (3,0)--(3,1);
\draw (3,0)--(4,0);
\draw (0,0) node[left]{$\Phi_{\vec{\alpha}_L} (\infty)$};
\draw (1,1) node[above]{$\Phi_{- n b \vec{\omega}_1+s\vec{\omega}_2} (1)$};
\draw (3,1) node[above]{$\Phi_{-b\vec{\omega}_1} (x)$};
\draw (4,0) node[right]{$\Phi_{\vec{\alpha}_R} (0)$};
\draw (2,0) node[below]{$M$};
\end{tikzpicture}
\end{aligned}
\end{equation}

\noindent The field $\Phi_{- b \vec{\omega}_1}$ is a fully degenerate field that is associated with the fundamental $\slthree$  representation, $\vec{\alpha}=\vec{\alpha}_{2111}$. The field $\Phi_{- n b \vec{\omega}_1+s\vec{\omega}_2}$ has one null-vector at level $n+1$, while the remaining fields  $\Phi_{\vec{\alpha}_L}$ and  $\Phi_{\vec{\alpha}_R}$ are general primaries: defining  $\vec{\alpha}_L \mydef a_{L_1}\vec{\omega}_1+a_{L_2}\vec{\omega}_2$ and  $\vec{\alpha}_R\mydef a_{R_1}\vec{\omega}_1+a_{R_2}\vec{\omega}_2$,  the four components $(a_{L_1}, a_{L_2},a_{R_1}, a_{R_2})$ are free parameters. 

The case $n=0$ was completely solved in \cite{fl07} where $\cB^{(0,s)}_M (x)$ was found to satisfy a Fuschian third-order equation. The case with $n=1$ and $s=-b$, corresponding  to a situation where the field at $x=1$ is fully-degenerate adjoint field, was considered in \cite{BEFS16} and is reviewed below. In \cite{BCES17}  the more general situation where $s$ is arbitrary was studied: $\cB^{(1,s)}_M (x)$ was shown to satisfy a sixth-order differential equation. 

We argue now, on the basis of the fusion rules and the properties \eqref{param} of the semi-degenerate fields, that the conformal block $\cB^{(n,s)}_{M} (x)$ satisfies a $3(n+1)$-th order Fuchsian ODE and that the corresponding $\iota=2$, i.e the system is rigid.

\paragraph{The order and the local exponents of the differential equation.}
\label{dimorigid}
Let us consider first the $s$-channel, i.e. the series expansion around $x=0$ of $\cB^{(n,s)}_M(x)$. From the fusion rules 
$\Phi_{-b \vec{\omega}_1} (x)\Phi_{R}(0)$, see (\ref{w3: fuprod}), we have three possibilities:
\begin{eqnarray}
\underline{\text{$s$-channel}}&& \nonumber \\
\label{fusion1}\text{Fusion 1:}\; \quad 
\vec{\alpha}_{M_1}^{(s)} &=& \vec{\alpha}_R- b \,     \vec{\omega}_1 \\
\label{fusion2}\text{Fusion 2:}\; \quad 
      \vec{\alpha}_{M_2}^{(s)} &=& \vec{\alpha}_R+ b \, \ll \vec{\omega}_1 - \vec{\omega}_2 \rr   \\
\label{fusion3}\text{Fusion 3:}\;\quad
\vec{\alpha}_{M_3}^{(s)} &=& \vec{\alpha}_R+ b \,     \vec{\omega}_2  
\end{eqnarray}

\noindent The above three channels determines the local exponents at $z=0$:
\begin{equation}
\cB^{(n,s)}_{M_i}(x) = x^{\alpha_{M_i}}\left(1+a_1 x +a_2 x^2+\cdots+a_{n} x^{n}+O(x^{n+1})\right),\quad i=1,2,3,
\end{equation}
where $\alpha_{M_i} = -h_{R}-h_{-b \vec{\omega}_1}+h_{\vec{\alpha}_{M_i}^{(s)}}$. The coefficients $a_1, \cdots, a_{n}$ are not fixed. This can be deduced by considering  the matrix elements $\lambda^{(l)}_{\vec{\alpha}_{L},\vec{\alpha}_{M_i}}(-n b \vec{\omega}_1+s \vec{\omega}_2)$ that enter into the computation of the conformal block, see (\ref{cbconsidered}) and (\ref{cb.series.expansion}). From the (\ref{param}) it follows that these matrix elements are not fixed by the algebra  for  $l=1,2,\cdots,n$. This means that the ODE possesses $(n+1)$-dimensional subspace of solutions characterized by the same critical exponent. Taking into account all the three fusions, the differential equation is expected to be of order $3(n+1)$ with the local exponents  multiplicities $(n+1,n+1,n+1)$ at $z=0$. The same arguments can be applied for the structure of local exponents and multiplicity at $z=\infty$ leading to the same results. Let us now consider the behavior of  $\cB^{(n,s)}_M(z)$ at $z=1$ ($t$-channel).  We have the following fusions:
 \begin{eqnarray}
\underline{\text{$t$-channel}} &&  \nonumber \\
\text{Fusion 1:}\;\quad 
\vec{\alpha}_{M_1}^{(t)} &=& - b(n+1) \,     \vec{\omega}_1 +s \vec{\omega}_2 \\
\text{Fusion 2:}\; \quad 
      \vec{\alpha}_{M_2}^{(t)} &=& - b(n-1) \,     \vec{\omega}_1 +(s-b) \vec{\omega}_2    \\
\text{Fusion 3:}\;\quad
\vec{\alpha}_{M_{3}}^{(t)} &=& - b n \,     \vec{\omega}_1 +(s+b) \vec{\omega}_2  
\end{eqnarray}
\noindent The above three channels determine the local exponents at $z=1$: 
\begin{equation}
\label{texpa}
\cB^{(n,s)}_{M_i}(x) = (x-1)^{\beta_{M_i}}\left(1+b_1 x +b_2 x^2+\cdots+b_{n+1} (x-1)^{n+1}+O((x-1)^{n+2})\right),\; i=1,2,3,
\end{equation}
where $\beta_{M_i} = -h_{-b\vec{\omega}_1}-h_{-n b \vec{\omega}_1+s \vec{\omega}_2}+h_{\vec{\alpha}_{M_i}^{(t)}}$.
At the $l$-level in  the  $t$-channel expansion (\ref{cb.series.expansion}) the matrix elements $\lambda^{(l)}_{\vec{\alpha}_{L},\vec{\alpha}_{R}}(\vec{\alpha}^{(t)}_{M_i})$, $i=1,2,3$,  will appear.  The internal channels $\vec{\alpha}_{M_i}$ are degenerate at level $n+2$ , $n$ and $n+1$ respectively for $i=1,2,3$ and therefore the matrix elements $\lambda^{(l)}_{\vec{\alpha}_{L},\vec{\alpha}_{R}}(\vec{\alpha}^{(t)}_{M_i})$ are undetermined for $l=1,\dots,n$ in the first fusion channel ($i=1$), for  $l=1,\dots,n-2$ in the second  fusion channel ($i=2$) and for $l=1,\dots,n-1$ in the third one ($i=3$). Correspondingly, the coefficients $b_j$ in (\ref{texpa}) are not fixed by the ODE for $j=1,\dots, n+1$, for $j=1,\dots, n-1$, and 
for $j=1,\cdots, n$ in the first, second and third fusion channel respectively. 
We arrive at the conclusion that $\cB^{(n,s)}_{M_i}(x)$ satisfies a Fuchsian ordinary differential
equation of rank $3(n+1)$ with multiplicities of local exponents $(n+1,n+1,n+1)$
at $x=0$ and $x=\infty$ and $(n+2,n+1,n)$ at $x=1$.
Symbolically, we denote these multiplicities as
$$
 ((n+1,n+1,n+1),
 (n+2,n+1,n),
 (n+1,n+1,n+1))\;.
$$

\section{Differential equations in CFT}
\label{diffeq}

In this article, we shall show how the Katz theory works for the study of differential equations
arising in CFT.
For the purpose, here we give several differential equations in CFT, the great part of which was already solved.
In the next sections, we shall derive these differential equations and their integral representations
of solutions by applying the Katz theory.

\subsection{Second order Virasoro ODE}
\label{Second-order}

We start from the simplest example of ODE appearing in CFT.
Consider the case of the Virasoro algebra $\mathfrak{V}_c$ with central charge 
$c=1+6(b+b^{-1})^2$. 
It is convenient to parametrize the conformal dimension $h$ of a primary field by its 
charge $a$, $h=a(b+b^{-1}-a)$. 
A primary field $\Phi_{a}$ with $a=-b/2$ (and dimension $h=-3/4 b^2-1$) is associate to a representation that is degenerate at second level \cite{rib14}.

We are interested here in the  4-point Virasoro conformal block $\cG_{{a}_M}(x)$ associated to the following diagram:
\begin{equation}
\begin{aligned}
\label{cbvir}
\begin{tikzpicture}
[line width=1.2pt]
\draw (-2.5,.5)node[left]{$\cG_{a_M} (x)$};
\draw (-2.5,.5) node[right]{$\mydef$};
\draw (0,0)--(1,0);
\draw (1,0)--(1,1);
\draw (1,0)--(3,0);
\draw (3,0)--(3,1);
\draw (3,0)--(4,0);
\draw (0,0) node[left]{$\Phi_{a_L} (\infty)$};
\draw (1,1) node[above]{$\Phi_{a_3} (1)$};
\draw (3,1) node[above]{$\Phi_{-b/2} (x)$};
\draw (4,0) node[right]{$\Phi_{a_R} (0)$};
\draw (2,0) node[below]{$a_M$};
\end{tikzpicture}
\end{aligned}
\end{equation}
Note that, as we said in the introduction, the Virasoro conformal block is completely specified once the dimensions of the external and internal  fields are given. 
That's why we can set $M=a_M$. 
By doing a gauge transform $\cG_{a_M}(x)= x^{a_R b}(1-x)^{a_3 b} G_{a_M}(x)$, 
one obtains
\begin{equation}
\label{rank2}
 x(1-x)\frac{d^2}{d x^2}
G_{a_M}(x)+\left[\gamma-(\alpha+\beta+1)x\right]\frac{d}{d x}
G_{a_M}(x)-\alpha\beta\; G_{a_M}(x)=0\;,
\end{equation}
where
\begin{align}
  \alpha= -1+ b (a_L+a_R+a_3)-\frac{3}{2}b^2\;,\quad  \beta= b (a_R+a_3-a_L)-\frac{1}{2}b^2\;, \quad  \gamma=2 b a_R-b^2\;.
 \end{align}
This differential equation (\ref{rank2}) is the Gauss (hypergeometric) equation,
which has the Riemann scheme
\begin{equation}
\label{Riem_not_2}
\begin{Bmatrix}  
x=0& & x=1 & & x=\infty  \\ 
0   & &  0   & & \alpha   \\
1-\gamma  & &  \gamma-\alpha-\beta   & & \beta   \\
\end{Bmatrix}.
\end{equation}
We specify a basis of solutions
\begin{equation}
\label{basis02}
 \begin{aligned}
  G_{a_R-b/2}(x) &= {}_2 F_{1}(\alpha,\beta;\gamma;x)=1+O(x)\;,\\
  G_{a_R+b/2}(x) &= x^{1-\gamma}{}_2 F_{1}(\alpha-\gamma+1,\beta-\gamma+1;2-\gamma;x)
  =x^{1-\gamma}(1+O(x))\;,
 \end{aligned}
\end{equation}
which are local solutions at $x=0$ of exponent $0$ and $1-\gamma$
with the initial term $1$ in the Taylor expansions.
The above solutions describe respectively the conformal blocks with the fusion channel 
$\Phi_{-b/2}\times\Phi_{a_R}(0)\to \Phi_{a_R-b/2}$ and  
$\Phi_{-b/2}\times\Phi_{a_R}(0)\to \Phi_{a_R+b/2}$.

\bigskip
The study of the global behavior of solutions of the Gauss equation (\ref{rank2}) allows to get the ratio of the structure constants for the conformal blocks. This method can be found in textbooks, see for instance chapter 9 of \cite{fms97}. In the following, we review in detail this study: it will make our notations more familiar to the reader and it will illustrate what we are going to do on more complicated case later in this article.

It is known that the Gauss equation (\ref{rank2}) has an integral representation of solutions:
\begin{equation}
\label{GaussInt}
 y(x)=\int_{\Delta}t^{\beta-\gamma}(t-1)^{\gamma-\alpha-1}(t-x)^{-\beta}\,dt\;.
\end{equation}
We assume that $x$ is a real number in the interval $(0,1)$.
By taking $\Delta=\Delta_{01}=\{1<t<\infty\}$ and $\Delta=\Delta_{02}=\{0<t<x\}$,
we get another basis of local solutions at $x=0$:
\begin{equation}
\label{GaussBasis2}
 \begin{aligned}
  y_{01}(x)&=\int_1^{\infty}t^{\beta-\gamma}(t-1)^{\gamma-\alpha-1}(t-x)^{-\beta}\,dt\;,\\
  y_{02}(x)&=\int_0^xt^{\beta-\gamma}(t-1)^{\gamma-\alpha-1}(t-x)^{-\beta}\,dt\;.
 \end{aligned}
\end{equation}
We see that $y_{01}(x)$ (res. $y_{02}(x)$) has exponent $0$ (resp. $1-\gamma$) at $x=0$.
Then we have a relation between two bases $(G_{a_R-b/2}(x),G_{a_R+b/2}(x))$ and
$(y_{01}(x),y_{02}(x))$ as
\begin{equation}
\label{YandG}
 y_{01}(x)=\mathcal{N}_1G_{a_R-b/2}(x)\;,\
 y_{02}(x)=\mathcal{N}_2G_{a_R+b/2}(x)\;,
\end{equation}
where
$$
 \begin{aligned}
  \mathcal{N}_1&=\int_0^1s^{\alpha-1}(1-s)^{\gamma-\alpha-1}\,ds
  =
  \frac{\Gamma(\alpha)\Gamma(\gamma-\alpha)}{\Gamma(\gamma)}\;,\\
  \mathcal{N}_2&=e^{\pi\sqrt{-1}(\gamma-\alpha-\beta-1)}
  \int_0^1s^{\beta-\gamma}(1-s)^{-\beta}\,ds
  =
  e^{\pi\sqrt{-1}(\gamma-\alpha-\beta-1)}
  \frac{\Gamma(\beta-\gamma+1)\Gamma(1-\beta)}{\Gamma(2-\gamma)}\;.
 \end{aligned}
$$
Note that in the integral on $\Delta_{02}=\{0<t<x\}$
we take $\arg(t-1)=\arg(t-x)=\pi$.

We also get a basis of local solutions at $x=1$ by taking
$\Delta=\Delta_{11}=\{-\infty<t<0\}$ and $\Delta=\Delta_{12}=\{x<t<1\}$:
\begin{equation}
\label{GaussBasis3}
 \begin{aligned}
  y_{11}(x)&=\int_{-\infty}^0t^{\beta-\gamma}(t-1)^{\gamma-\alpha-1}(t-x)^{-\beta}\,dt\;,\\
  y_{12}(x)&=\int_x^1t^{\beta-\gamma}(t-1)^{\gamma-\alpha-1}(t-x)^{-\beta}\,dt\;.
 \end{aligned}
\end{equation}
By the local analysis of these integrals,
we get
\begin{equation}
\label{GaussX=1}
 \begin{aligned}
  y_{11}(x)&=e^{\pi\sqrt{-1}(-\alpha-1)}\frac{\Gamma(\alpha)\Gamma(\beta-\gamma+1)}
  {\Gamma(\alpha+\beta-\gamma+1)}(1+O(x-1))\;,\\
  y_{12}(x)&=e^{\pi\sqrt{-1}(\gamma-\alpha-1)}
  \frac{\Gamma(1-\beta)\Gamma(\gamma-\alpha)}{\Gamma(\gamma-\alpha-\beta+1)}
  (1-x)^{\gamma-\alpha-\beta}(1+O(x-1))\;.
 \end{aligned}
\end{equation}

On the other hand,
we can derive the linear relations among the twisted cycles $\Delta_{ij}$ $(i=0,1;j=1,2)$ from
Cauchy's theorem.
Explicitly, we get
$$
 \begin{aligned}
  \Delta_{11}+\Delta_{02}+\Delta_{12}+\Delta_{01}&=0\;,\\
  \Delta_{11}+e_1\Delta_{02}+e_1e_2\Delta_{12}+e_1e_2e_3\Delta_{01}&=0\;,
 \end{aligned}
$$
where we set
$$
 e_1=e^{2\pi\sqrt{-1}(\beta-\gamma)}\;,\
 e_2=e^{2\pi\sqrt{-1}(-\beta)}\;,\
 e_3=e^{2\pi\sqrt{-1}(\gamma-\alpha)}\;.
$$
These relations can be regarded as linear relations among $y_{ij}(x)$ $(i=0,1;j=1,2)$,
and we solve them to get
$$
 \begin{aligned}
  y_{01}(x)&=\frac{1-e_1}{e_1(1-e_2e_3)}y_{11}(x)+\frac{e_2-1}{1-e_2e_3}y_{12}(x)\;,\\
  y_{02}(x)&=\frac{e_1e_2e_3-1}{e_1(1-e_2e_3)}y_{11}(x)
  +\frac{e_2(e_3-1)}{1-e_2e_3}y_{12}(x)\;.
 \end{aligned}
$$
Namely, we get the connection relation
\begin{equation}
\label{GaussConnection}
 (y_{01}(x),y_{02}(x))=(y_{11}(x),y_{12}(x))C_{10}\;,\quad\
 C_{10}=
  \begin{pmatrix}\frac{1-e_1}{e_1(1-e_2e_3)}&\frac{e_1e_2e_3-1}{e_1(1-e_2e_3)}\\
  \frac{e_2-1}{1-e_2e_3}&\frac{e_2(e_3-1)}{1-e_2e_3}
  \end{pmatrix}.
\end{equation}

Now we compute the monodromy with respect to the basis $(y_{01}(x),y_{02}(x))$.
Let $\gamma_0$ and $\gamma_1$ be loops with base point $x=1/2$ which encircle $x=0$ and $x=1$,
respectively, once in the positive direction.
We denote by ${\gamma_j}_*$ the analytic continuation along $\gamma_j$ $(j=0,1)$.
From the local behaviors (\ref{basis02}), (\ref{YandG}) and (\ref{GaussX=1}),
we readily obtain
$$
 {\gamma_0}_*(y_{01}(x),y_{02}(x))=(y_{01}(x),y_{02}(x))M_0\;,\quad\
 {\gamma_1}_*(y_{11}(x),y_{12}(x))=(y_{11}(x),y_{12}(x))\tilde M_1\;,
$$
where
$$
 M_0=\begin{pmatrix}1&\\ &e^{2\pi\sqrt{-1}(-\gamma)}\end{pmatrix},\
\quad \tilde M_1=\begin{pmatrix}1&\\ &e^{2\pi\sqrt{-1}(\gamma-\alpha-\beta)}\end{pmatrix}.
$$
Then, by combining with the connection relation (\ref{GaussConnection}),
we obtain the monodromy representation
$$
 \begin{aligned}
  {\gamma_0}_*(y_{01}(x),y_{02}(x))&=(y_{01}(x),y_{02}(x))M_0\;,\\
  {\gamma_1}_*(y_{01}(x),y_{02}(x))&=(y_{01}(x),y_{02}(x))M_1\;,
 \end{aligned}
$$
with
$$
 M_1=C_{10}^{-1}\tilde M_1C_{10}
 =\begin{pmatrix}\frac{e_2(1-e_1+e_1e_3-e_1e_2e_3)}{1-e_1e_2}&
 \frac{e_2(1-e_3)(1-e_1e_2e_3)}{e_1e_2-1}\\
 \frac{(1-e_1)(1-e_2)}{e_1e_2-1}&
 \frac{1-e_2+e_2e_3-e_1e_2e_3}{1-e_1e_2}
 \end{pmatrix}.
$$
The matrices $M_0,M_1$ are generators of the monodromy group of the equation
(\ref{rank2}) with respect to the basis $(y_{01}(x),y_{02}(x))$.

From now on we assume that the parameters $\alpha,\beta,\gamma$ are real numbers.
We compute the monodromy invariant Hermitian form.
A Hermitian form is monodromy invariant if it is given by a Hermitian matrix $H$
satisfying
$$
 \bar MHM^T=H
$$
for any matrix $M$ in the monodromy group.
By the conditions
$$
 \bar M_0HM_0^T=H\;,\
 \bar M_1HM_1^T=H\;,
$$
we get
$$
 H=h\begin{pmatrix}1&\\ &\frac{(1-e_1)(1-e_2)e_3}{(1-e_3)(1-e_1e_2e_3)}\end{pmatrix}
$$
with any real number $h$.
Therefore the monodromy invariant Hermitian form exists uniquely up to multiplication
by $\mathbb{R}^{\times}$.

As discussed in the introduction, the monodromy invariant Hermitian form $H$ provides
the factorization of the correlation function:
$$
 \langle\Phi_{a_L}(\infty)
 \Phi_{a_3}(1)
 \Phi_{-b/2}(x)
 \Phi_{a_R}(0)\rangle
 =\mathcal{Y}_0^*(x)H\mathcal{Y}_0(x)^T\;,
$$
where we set
$$
 \mathcal{Y}_0(x)=(y_{01}(x),y_{02}(x))\;.
$$
By taking the expansion of the right hand side at $x=0$, we get
$$
 \begin{aligned}
 &\langle\Phi_{a_L}(\infty)
 \Phi_{a_3}(1)
 \Phi_{-b/2}(x)
 \Phi_{a_R}(0)\rangle\\
 &\qquad\qquad
 =h\left(|\mathcal{N}_1|^2(1+O(x))
 +x^{2(1-\gamma)}
 |\mathcal{N}_2|^2\frac{(1-e_1)(1-e_2)e_3}{(1-e_3)(1-e_1e_2e_3)}(1+O(x))\right),
 \end{aligned}
$$
from which we derive the ratio of the structure constant
\begin{equation}
\label{GaussSC}
 \begin{aligned}
  &\frac{C(a_{R}+b/2,-b/2, a_{R})C(a_L, a_{3}, a_{R}+b/2) }
  {C(a_{R}-b/2,-b/2, a_{R})C(a_L, a_{3}, a_{R}-b/2)}\\
  &\quad
  =\frac{|\mathcal{N}_2|^2\frac{(1-e_1)(1-e_2)e_3}{(1-e_3)(1-e_1e_2e_3)}}{|\mathcal{N}_1|^2}\\
  &\quad
  =
  \frac{\Gamma(1-\alpha)\Gamma(1-\beta)\Gamma(1-\gamma+\alpha)
  \Gamma(\gamma)^2}
  {\Gamma(\alpha)\Gamma(\beta)\Gamma(\gamma-\alpha)
  \Gamma(2-\gamma)^2}\\
  &\quad
  =
  \frac{\Gamma(2-b(a_L+a_R+a_3)+\frac{3}{2}b^2)
  \Gamma(1+b(a_L-a_R-a_3)+\frac{1}{2}b^2)}
  {\Gamma(-1+b(a_L+a_R+a_3)-\frac{3}{2}b^2)
  \Gamma(b(a_R+a_3-a_L)-\frac{1}{2}b^2)}\\
  &\qquad
  \times
  \frac{\Gamma(b(a_L-a_R+a_3)-\frac{1}{2}b^2)
  \Gamma(2ba_R-b^2)^2}
  {\Gamma(1-b(a_L-a_R+a_3)+\frac{1}{2}b^2)
  \Gamma(2-2ba_R+b^2)^2}\;.
 \end{aligned}
\end{equation}

\subsection{The case of $\cB_{M}^{(0,s)}(x)$: the third-order differential equation}
\label{rank3sol}

We return to the $\cW_3$ algebra introduced in Section 2 and we focus on the $\cW_3$ conformal block $\cB_{M}^{(0,s)}(x)$, defined in (\ref{cb_sem_deg}). The conformal block $\cB_{M}^{(0,s)}(x)$ is built by matrix elements involving one fully-degenerate operator and one semi-degenerate operator at level one. Therefore, following the discussion in (\ref{degeneratefields}), there are no additional internal parameters. We can therefore set $M=\vec{\alpha}_M$.  After operating a gauge transform:
\begin{equation}
\cB_{\vec{\alpha}_M}^{(0,s)}(x)= x^{\frac{1}{3}(2 a_{R_1} + a_{R_2}) b}(1-x)^{\frac{b s}{3}}G_{\vec{\alpha}_M}(x)\;,
\end{equation}
the function $G_{\vec{\alpha}_M}(x)$ satisfies
the generalized hypergeometric equation \cite{fl07c}:
\begin{multline}
\label{Pochgamer}
  \left[x\left(x\frac{d}{dx}+p_1\right)
  \left(x\frac{d}{dx}+p_2\right)
  \left(x\frac{d}{dx}+p_3\right)-\right.\\-\left.
  \left(x\frac{d}{dx}+q_1-1\right)
  \left(x\frac{d}{dx}+q_2-1\right)x\frac{d}{dx}\right]G_{\vec{\alpha}_M}(x)=0\;,\;
\end{multline}
where
\begin{equation}
    p_k=\frac{b s}{3}-\frac{2}{3}b^2+b(\vec{\alpha}_R-Q \vec{\rho},\vec{h}_1)+b(\vec{\alpha}_L-Q\vec{\rho},\vec{h}_k)\;, \quad k=1,2,3
\end{equation}  
and
\begin{equation}
  \begin{aligned}
   q_1=1+b(\vec{\alpha}_R-Q\vec{\rho},\vec{e}_1)\;, \quad q_2=1+b(\vec{\alpha}_R-Q\vec{\rho},\vec{e}_1+\vec{e}_2)\;.
  \end{aligned}
\end{equation}
The three dimensional space of solutions is spanned by the basis of functions:
\begin{equation}
\label{basis3}
 \begin{aligned}
& G_{\vec{\alpha}_R-b\vec{\omega}_1}(x)=
  {}_3F_2\left[p_1,p_2,p_3;q_1,q_2;x\right]\;, \\
& G_{\vec{\alpha}_R+b\vec{\omega}_1-b\vec{\omega}_2}(x)=
  x^{1-q_1}{}_3F_2\left[1- q_1+p_1,1-q_1+p_2,1-q_1+p_3;
 2-  q_1,1-q_1+q_2;x\right]\;, \\
 &G_{\vec{\alpha}_R-b\vec{\omega}_2}(x)=
  x^{1-q_2}{}_3F_2\left[1-q_2+p_1,1-q_2+p_2,1-q_2+p_3;
 1-q_2+q_1,2-q_2;x\right] \;,
 \end{aligned}
\end{equation}
that correspond respectively to the fusions (\ref{fusion1}), (\ref{fusion2}) and (\ref{fusion3}). 
This ODE has the following Riemann scheme:
\begin{equation}
 \label{Riem_not_3}
  \begin{Bmatrix}  
   x=0& & x=1 & & x=\infty  \\ 
   0   & &  0   & & p_1   \\
   1-q_1  & &  1   & & p_2   \\
   1-q_2   & &  -p_1-p_2-p_3+q_1+q_2  & & p_3  
  \end{Bmatrix}.
\end{equation}
As in the case of Gauss hypergeometric equation,
it is also known that the differential equation (\ref{Pochgamer}) has an integral representation of
solutions:
$$
 y(x)=\int_{\Delta}t_1^{p_2-q_2}(t_1-1)^{q_2-p_3-1}(t_1-t_2)^{q_1-p_2-1}
 t_2^{p_1-q_1}(t_2-x)^{-p_1}
 \,dt_1\wedge dt_2\;.
$$
Then, by using this integral representation,
we can compute the monodromy and hence the monodromy invariant Hermitian form.
Therefore, in a similar manner as Section~\ref{Second-order},
we can derive the ratio of the structure constants.

\subsection{The conformal block $\cB^{(1,-b)}_{M}(x)$: fourth-order differential equation}

The conformal block $\cB^{(1,-b)}_{M}(x)$, defined in (\ref{cbconsidered}),
satisfies the fourth-order Fuchsian differential equation with the Riemann scheme
\cite{BEFS16}:
\begin{equation}
\label{rank4RS}
 \left\{
  \begin{matrix}
   x=0&x=1&x=\infty\\
   \alpha_1&\beta_1&\gamma_1\\
   \alpha_2&\beta_2&\gamma_2\\
   \alpha_3&\beta_3&\gamma_3\\
   \alpha_3+1&\beta_3+1&\gamma_3+1
  \end{matrix}
 \right\},
\end{equation}
where
\begin{equation}
\label{4.18}
 \begin{aligned}
  \alpha_1&=-\frac{a_{R_1}}{3}b+\frac{a_{R_2}}{3}b+b^2+1\;,\
  \alpha_2=2-\frac{a_{R_1}}{3}b-\frac{2a_{R_2}}{3}b+2b^2\;,\
  \alpha_3=\frac{2a_{R_1}}{3}b+\frac{a_{R]2}}{3}b\;,\\
  \beta_1&=1+b^2\;,\
  \beta_2=2+3b^2\;,\
  \beta_3=-b^2\;,\\
  \gamma_1&=-2+\frac{a_{R_1}}{3}b+\frac{2a_{R_2}}{3}b-3b^2\;,\
  \gamma_2=-1+\frac{a_{R_1}}{3}b-\frac{a_{R_2}}{3}b-2b^2\;,\
  \gamma_3=-\frac{2a_{R_1}}{3}b-\frac{a_{R_2}}{3}b\;.
 \end{aligned}
\end{equation}
In \cite{BEFS16}, the explicit form of the fourth order differential equation is given.
Also an integral representation of solutions, the monodromy group
and the monodromy invariant Hermitian form are obtained,
and then the structure constants are derived.

\subsection{The conformal block $\cB^{(1,s)}_{M}(x)$: sixth-order differential equation}
The conformal block $\cB^{(1,s)}_{M}(x)$, where at position $x=1$ instead of a fully-degenerate adjoint field there  a second-level semi-degenerate field, has been shown to satisfy a sixth-order Fuchsian differential equation with the following  Riemann scheme \cite{BCES17}:
\begin{equation}
\label{Riem_rank6}
\begin{Bmatrix}  
x=0& & x=1 & & x=\infty  \\ 
\alpha_1   & &  \beta_1   & & \gamma_1   \\
\alpha_1 +1  & &  \beta_1 +1   & & \gamma_1 +1   \\
\alpha_2   & &  \beta_1+2  & & \gamma_2   \\
\alpha_2 +1 & &  \beta_2 & & \gamma_2 +1  \\
\alpha_3 && \beta_3 && \gamma_3 \\
\alpha_3 +1 && \beta_3 +1  && \gamma_3 +1
\end{Bmatrix}
\end{equation}

\noindent with

\begin{eqnarray}
\alpha_1 &=&\frac{2 a_{R_1} b + a_{R_2} b}{3}\;,
\quad
\alpha_2 =\frac{3 - a_{R_1} b + a_{R_2} b + 3 b^2}{3}\;,  
\quad   
\alpha_3 =\frac{6 - a_{R_1} b - 2 a_{R_2} b + 6 b^2}{3}\;,  
\nonumber \\
\beta_1 &=& \frac{-2 b^2 + b s}{3}\;,
\quad \beta_2 =\frac{3 + 4 b^2 + b s}{3}\;, 
\quad
\beta_3 = \frac{6 + 7 b^2 - 2 b s}{3}\;,\nonumber \\
\label{Riem_param_6}
\gamma_1 &=& \frac{a_{L_1}b - a_{L_2}b - 5 b^2-3}{3}\;,\,
\gamma_2 = \frac{-6 + a_{L_1} b + 2 a_{L_2} b - 8 b^2}{3}\;,\,
\gamma_3 = \frac{-2 a_{L_1} b - a_{L_2} b - 2 b^2}{3}\;.  \nonumber \\
&&
\end{eqnarray} 
In \cite{BCES17}, the solutions and their monodromy properties were not found. 
We provide below a complete set of solutions of the sixth order Fuchsian equation by applying the Katz theory. 

\section{Katz theory on rigid local systems}
\label{Katz-theory}

In this section we briefly review the Katz theory on rigid local systems \cite{katz1996rigid}.  
More detailed consideration can be found,  e.g., in~\cite{DR1, DR2}.

\subsection{Katz theory for Fuchsian systems}

Let us consider a Fuchsian system
\begin{equation}
\label{FuchsSystem}
 \frac{dY}{dx}=\ll\sum_{j=1}^p\frac{A_j}{x-a_j}\rr Y\;,
\end{equation}
where $a_1,a_2,\dots,a_p$ are distinct points in $\mathbb{C}$ and $A_1,A_2,\dots,A_p$
are constant $n\times n$-matrices. 
We introduce the residue matrix at $x=\infty$ by
$$
 A_0=-\sum_{j=1}^pA_j\;.
$$
Then obviously we have $\sum_{j=0}^p{\rm tr}A_j=0$,
which implies that the sum of all eigenvalues of all residue matrices $A_j$ is zero.
This equality corresponds to the Fuchs relation for scalar Fuchsian ordinary differential equations.
(We recall the Fuchs relation in Section 4.2.)
We assume that every $A_j$ $(0\leq j\leq p)$ is non-resonant, that is there is no integral difference among distinct eigenvalues. Nevertheless, multiplicities of the eigenvalues are allowed to be present.

The tuple $(A_0,A_1,\dots,A_p)$ is called {\it irreducible}
if there is no non-trivial common invariant
subspace in $\mathbb{C}^n$, and is called {\it rigid} if the equivalence class
$[(A_0,A_1,\dots,A_p)]$ is uniquely determined by the conjugacy classes
$[A_0],[A_1],\dots,[A_p]$,
where the equivalence relation is defined by
$$
 \begin{aligned}
 &(A_0,A_1,\dots,A_p)\sim(B_0,B_1,\dots,B_p)\\
 &\quad\Leftrightarrow
 A_j=PB_jP^{-1}\quad(0\leq j\leq p)\quad(\exists P\in{\rm GL}(n,\mathbb{C}))\;.
 \end{aligned}
$$
In other words, the tuple $(A_0,A_1,\dots,A_p)$ is rigid
if, for any tuple $(B_0,B_1,\dots, B_p)$ with $B_0 +\cdots + B_p = O$ satisfying
$B_j = P_jA_jP_j^{-1}$ with some non-singular matrix $P_j$ for $j = 0, \dots, p$, there exists a
non-singular matrix $P$ such that $B_j = P A_j P^{-1}$ for $j = 0,\cdots ,p$.

There is a simple criterion for the rigidity.
Katz defined the {\it index of rigidity} by
\begin{equation}
\label{iota}
 \iota=(1-p)n^2+\sum_{j=0}^p\dim Z(A_j)\;,
\end{equation}
where $Z(A_j)$ denotes the centralizer of $A_j$.
Then it is shown that, if the tuple $(A_0,A_1,\dots,A_p)$ is irreducible, $\iota\leq2$ holds.
In this case, the tuple is rigid if and only if $\iota=2$.
We call the Fuchsian system (\ref{FuchsSystem}) rigid if the tuple of the residue matrices is rigid.

Now we define two basic operations:

{\bf i.} The {\it addition} with the parameter $(\alpha_1,\alpha_2,\dots,\alpha_p)\in\mathbb{C}^p$ is
the operation
$$
 (A_1,A_2,\dots,A_p)\to(A_1+\alpha_1,A_2+\alpha_2,\dots,A_p+\alpha_p)\;,
$$
where we use the notation $A+\alpha=A+\alpha I$.
This operation corresponds to the gauge transform
\begin{equation}
\label{gauge}
 Y(x)\to Y(x)\prod_{j=1}^p(x-a_j)^{\alpha_j}
\end{equation}
for the system (\ref{FuchsSystem}).

{\bf ii.} The second operation is called the {\it middle convolution}.
Let $\lambda$ be a complex number.
We first define the operation
\begin{equation}
 (A_1,A_2,\dots,A_p)\to(B_1,B_2,\dots,B_p)
\label{mc1}
\end{equation}
by
\begin{equation}
\label{mc2}
 B_j=\sum_{k=1}^pE_{jk}\otimes(A_k+\delta_{jk}\lambda)
 \quad(1\leq j\leq p)\;,
\end{equation}
where $E_{jk}$ denotes the $p\times p$-matrix with the only non-zero entry 1
at $(j,k)$-th position $(1\leq j,k\leq p)$.
Let $\mathfrak{k}$ and $\mathfrak{l}$ be the subspaces of $\mathbb{C}^{pn}$ defined by
$$
 \begin{aligned}
  \mathfrak{k}&=\left\{
   \begin{pmatrix}v_1\\ \vdots\\ v_p\end{pmatrix}\,;\,
   v_j\in{\rm Ker} A_j\ (1\leq j\leq p)\right\}\;,\\
  \mathfrak{l}&={\rm Ker}(B_1+B_2+\cdots+B_p)\;.
 \end{aligned}
$$
It is easy to see  that $\mathfrak{k}$ and $\mathfrak{l}$ 
are invariant subspaces of $\mathbb{C}^{pn}$
for $(B_1,B_2,\dots,B_p)$.
Then $(B_1,B_2,\dots,B_p)$ induces the action 
$(\bar B_1,\bar B_2,\dots,\bar B_p)$ on the quotient space 
$\mathbb{C}^{pn}/(\mathfrak{k}+\mathfrak{l})$.
The operation
$$
 (A_1,A_2,\dots,A_p)\to(\bar B_1,\bar B_2,\dots,\bar B_p)
$$
is called the middle convolution with parameter $\lambda$, and is denoted by $mc_{\lambda}$.
The middle convolution corresponds to
the composition of the Riemann-Liouville transform (or Euler transform)
\begin{equation}
\label{RiemannLiouville}
 Y(x)\to\int_{\Delta}Y(t)(t-x)^{\lambda}\,\vec\eta
\end{equation}
with
$$
 \vec\eta=\left(\frac{dt}{t-a_1},\frac{dt}{t-a_2},\dots,\frac{dt}{t-a_p}\right)^T
$$
and the linear transform which realizes the quotient by $\mathfrak{k}+\mathfrak{l}$.

The middle convolution possesses several basic properties.
Under some generic condition, the additivity
\begin{equation}
\label{mcAddtive}
 mc_0={\rm id.},\
 mc_{\lambda}\circ mc_{\mu}=mc_{\lambda+\mu}
\end{equation}
holds.
Moreover, the middle convolution keeps the index of rigidity and the irreducibility.
We sometimes call $mc_{\lambda}$ an additive middle convolution.

Katz showed a remarkable assertion:
Any irreducible rigid tuple is obtained from a tuple of scalars by a finite iteration
of additions and middle convolutions.
This result implies that rigid Fuchsian system (\ref{FuchsSystem}) can be obtained from a rank one equation
\begin{equation}
\label{rank1}
 \frac{dy}{dx}=\ll\sum_{j=1}^p\frac{\alpha_j}{x-a_j}\rr y
\end{equation}
with $\alpha_j\in\mathbb{C}$ by a finite iteration of gauge transforms (\ref{gauge})
and Riemann-Liouville transforms (\ref{RiemannLiouville}).
Hence we have an integral representation of solutions for any rigid Fuchsian system (cf. \cite{HH}).

By the Riemann-Hilbert correspondence, the above results can be translated to monodromies.
Let
$$
 \rho:\pi_1(\mathbb{P}^1\setminus\{a_0,a_1,\dots,a_p\},b)\to{\rm GL}(n,\mathbb{C})
$$
be an anti-representation.
We can regard $\rho$ as a local system over $\mathbb{P}^1\setminus\{a_0,a_1,\dots,a_p\}$,
and as the monodromy representation of the system (\ref{FuchsSystem}).
The fundamental group is presented
as
$$
 \pi_1(\mathbb{P}^1\setminus\{a_0,a_1,\dots,a_p\},b)
 =
 \langle\gamma_0,\gamma_1,\dots,\gamma_p\mid
 \gamma_0\gamma_1\cdots\gamma_p=1\rangle\;,
$$
where each $\gamma_j$ is a loop which encircles $a_j$ once in the positive direction
and does not encircle the other $a_k$'s.
We set $M_j=\rho(\gamma_j)$ for $0\leq j\leq p$.
Then we have $M_p\cdots M_1M_0=I$, and
$\rho$ is determined by the tuple $(M_0,M_1,\dots,M_p)$.
We call the conjugacy class $[M_j]$ the local monodromy at $a_j$.
The local system
$\rho$ is said to be rigid if it is uniquely determined up to isomorphisms
by the local monodromies $[M_0],[M_1],\dots,[M_p]$.
We have a similar criterion for the rigidity of local systems using the index of rigidity (\ref{iota})
with $A_j$ replaced by $M_j$.

Analogously to the case of Fuchsian systems,
we can define the multiplication and the middle convolution for the local system
$\rho=(M_0,M_1,\dots,M_p)$.
The multiplication with parameter $(\beta_1,\beta_2,\dots,\beta_p)\in(\mathbb{C}^{\times})^p$
is the operation
\begin{equation}
\label{add_mon}
 (M_1,M_2,\dots,M_p)\to(\beta_1M_1,\beta_2M_2,\dots,\beta_pM_p)\;.
\end{equation}
The definition of the (multiplicative) middle convolution is as follows.
Let $\mu$ be an element in $\mathbb{C}^{\times}$.
We first define the operation
\begin{equation}
\label{middle_mon1}
(M_1,M_2,\dots,M_p)\to(G_1,G_2,\dots,G_p)
\end{equation}
by
\begin{equation}
\label{middle_mon2}
 G_j=I_{pn}+\sum_{k=1}^{j-1}E_{jk}\otimes\mu(M_k-1)+E_{jj}\otimes(\mu M_j-1)
 +\sum_{k=j+1}^pE_{jk}\otimes(M_k-1)
 \quad(1\leq j\leq p)\;.
\end{equation}
Let $\mathcal{K}$ and $\mathcal{L}$ be the subspaces of $\mathbb{C}^{pn}$ defined by
$$
 \begin{aligned}
  \mathcal{K}&=\left\{\
   \begin{pmatrix}v_1\\ \vdots\\ v_p\end{pmatrix}\,;\,
   v_j\in{\rm Ker}(M_j-1)\ (1\leq j\leq p)\right\},\\
  \mathcal{L}&={\rm Ker}(G_1G_2\cdots G_p-1)\;.
 \end{aligned}
$$
It is easy to see that $\mathcal{K}$ and $\mathcal{L}$ are invariant subspaces of 
$\mathbb{C}^{pn}$ for $(G_1,G_2,\dots,G_p)$.
Then $(G_1,G_2,\dots,G_p)$ induces the action $(\bar G_1,\bar G_2,\dots,\bar G_p)$
on the quotient space $\mathbb{C}^{pn}/(\mathcal{K}+\mathcal{L})$.
The operation
$$
 (M_1,M_2,\dots,M_p)\to(\bar G_1,\bar G_2,\dots,\bar G_p)
$$
is said to be the {\it (multiplicative) middle convolution} with parameter $\mu$,
and is denoted by $MC_{\mu}$.
The multiplicative middle convolution $MC_{\mu}$ enjoys similar properties as $mc_{\lambda}$.

It is useful to look at spectral types.
For simplicity, we consider only semi-simple (i.e. diagonalizable) matrices.
The spectral type of a semi-simple matrix $A$ is the partition which describes the multiplicities
of the eigenvalues of $A$.
For a tuple $(A_0,A_1,\dots,A_p)$ of semi-simple matrices,
its spectral type is defined by the tuple of the spectral types of the entries.
For a tuple $(A_0,A_1,\dots,A_p)$ of the residue matrices of the Fuchsian system
(\ref{FuchsSystem}),
the addition does not change the spectral type.
It is shown that the spectral type changes in the middle convolution as follows.
For $j\neq0$, let $(m_{j0}^*,m_{j1},\dots,m_{jn_j})$ be the spectral type of $A_j$,
where the marked entry $m_{j0}^*$ denotes the multiplicity of the eigenvalue $0$.
Let $(m_{00}^*,m_{01},\dots,m_{jn_0})$ be the spectral type of $A_0$,
where the marked entry $m_{00}^*$ denotes the multiplicity of the eigenvalue $\lambda$
which is the parameter of the middle convolution.
Owing to this convention, $m_{j0}^*,m_{00}^*$ can take value $0$.
Set
\begin{equation}
\label{d}
 d=\sum_{j=0}^pm_{j0}^*-(p-1)n\;.
\end{equation}
Then in the middle convolution $mc_{\lambda}$,
the spectral type of each residue matrix $A_j$ changes as
\begin{equation}
\label{mc4}
  (m_{j0}^*,m_{j1},\dots,m_{jn_j})\to
  (m_{j0}^*-d,m_{j1},\dots,m_{jn_j})\quad(0\leq j\leq p)\;.
\end{equation}
We can also give the changes of the eigenvalues in the middle convolution as
$$
 \begin{aligned}
  \begin{pmatrix}0&\alpha_{j1}&\cdots&\alpha_{jn_j}\\
  m_{j0}^*&m_{j1}&\cdots&m_{jn_j}
  \end{pmatrix}
  &\to
  \begin{pmatrix}0&\alpha_{j1}+\lambda&\cdots&\alpha_{jn_j}+\lambda\\
  m_{j0}^*-d&m_{j1}&\cdots&m_{jn_j}
  \end{pmatrix}&&(j\neq0)\;,\\
  \begin{pmatrix}\lambda&\alpha_{01}&\cdots&\alpha_{0n_0}\\
  m_{00}^*&m_{01}&\cdots&m_{0n_0}
  \end{pmatrix}
  &\to
  \begin{pmatrix}-\lambda&\alpha_{01}-\lambda&\cdots&\alpha_{0n_0}-\lambda\\
  m_{00}^*-d&m_{01}&\cdots&m_{0n_0}
  \end{pmatrix}&&(j=0)\;.
 \end{aligned}
$$

By the description (\ref{mc4}) of the change of the spectral types,
we see how we can connect a rigid Fuchsian system to a rank one equation.
Let
$$
 ((m_{00},m_{01},\dots,m_{0n_0}),(m_{10},m_{11},\dots,m_{1n_1}),\dots,
 (m_{p0},m_{p1},\dots,m_{pn_0}))
$$
be the spectral type of the tuple $(A_0,A_1,\dots,A_p)$ of the residue matrices of the Fuchsian
system (\ref{FuchsSystem}).
We assume that, for each $j$, $m_{j0}$ is the maximum among $m_{j0},m_{j1},\dots,m_{jn_j}$.
We operate an addition so that the eigenvalue of multiplicity $m_{j0}$ becomes $0$ for each $j$.
Also we take as $\lambda$ the eigenvalue of $A_0$ of multiplicity $m_{00}$.
It is shown that, if the Fuchsian system (\ref{FuchsSystem}) is irreducible and rigid,
the integer $d$ defined by (\ref{d}) becomes positive.
Then, thanks to (\ref{mc4}),
we can reduce the rank of the Fuchsian system
by operating the middle convolution $mc_{\lambda}$ with parameter $\lambda$.
By iterating these two steps, we come to a rank one equation (\ref{rank1}).
Evidently the addition is an invertible operation,
and, owing to the additivity (\ref{mcAddtive}),
the middle convolution is also invertible.
Then by taking the inverse operations, we have a chain of additions and middle convolutions
starting from a rank one equation and arriving at the given rigid Fuchsian system.

To sum up,
the Katz theory for Fuchsian systems provides recursive ways of constructing
Fuchsian systems, integral representations of their solutions,
and their monodromy representations.
Moreover, if the Fuchsian system is rigid, which can be easily checked only by calculating
the index of rigidity,
then all the above quantities are obtained by a chain of additions and middle convolutions
starting from ones of a rank one equation.

\subsection{Katz theory for scalar Fuchsian equations}

We note that in \cite{oshima} a similar theory for scalar differential equations was developed, which may be also useful in some respects.
In this subsection,
we explain how to get the Riemann scheme and an integral representation of solutions
for scalar equations
when we operate Katz's basic operations.

We consider a scalar Fuchsian differential equation
\begin{equation}
\label{scalarN}
 y^{(n)}+p_1(x)y^{(n-1)}+\cdots+p_{n-1}(x)y'+p_n(x)y=0
\end{equation}
of order $n$.
We denote the singular points by $a_1,a_2,\dots,a_p$ and $a_0=\infty$.
The Riemann scheme is written as
\begin{equation}
\label{scalarRS}
 \left\{
  \begin{matrix}
   x=a_1&x=a_2&\cdots&x=a_p&x=\infty\\
   [\lambda_{10}]_{(m_{10})}&[\lambda_{20}]_{(m_{20})}&\cdots
    &[\lambda_{p0}]_{(m_{p0})}&[\lambda_{00}]_{(m_{00})}\\
   [\lambda_{11}]_{(m_{11})}&[\lambda_{21}]_{(m_{21})}&\cdots
    &[\lambda_{p1}]_{(m_{p1})}&[\lambda_{01}]_{(m_{01})}\\
   \vdots&\vdots&&\vdots&\vdots\\
   [\lambda_{1n_1}]_{(m_{1n_1})}&[\lambda_{2n_2}]_{(m_{2n_2})}&\cdots
    &[\lambda_{pn_p}]_{(m_{pn_p})}&[\lambda_{0n_0}]_{(m_{0n_0})}
  \end{matrix}
 \right\},
\end{equation}
where the symbol $[\lambda]_{(m)}$ means
$$
 [\lambda]_{(m)}
 =
 \begin{pmatrix}\lambda\\ \lambda+1\\ \vdots\\ \lambda+m-1\end{pmatrix},
$$
and moreover there is no logarithmic solution of exponent $\lambda$.
For each $j$, $(m_{j0},m_{j1},\dots,m_{jn_j})$ is the partition of $n$
describing the multiplicities of the characteristic exponent
$\lambda_{j0},\lambda_{j1},\dots,\lambda_{jn_j}$, respectively,
and is called the spectral type at $x=a_j$.
We also call the tuple of the spectral types
$$
 ((m_{00},m_{01},\dots,m_{0n_0}),(m_{10},m_{11},\dots,m_{1n_1}),\dots,
 (m_{p0},m_{p1},\dots,m_{pn_0}))
$$
the spectral type of the scalar equation (\ref{scalarN}).
For the characteristic exponents $\lambda_{j\nu}$, the Fuchs relation
\begin{equation}
\label{FuchsRelation}
 \text{The sum of all characteristic exponents}=\frac{n(n-1)(p-1)}{2}
\end{equation}
holds.
In terms of $\lambda_{j\nu}$ and $m_{j\nu}$, it is written as
$$
 \sum_{j=0}^p\sum_{\nu=0}^{n_j}\sum_{i=0}^{m_{j\nu}-1}(\lambda_{j\nu}+i)
 =\frac{n(n-1)(p-1)}{2}\;.
$$
The index of rigidity is also defined for a scalar differential equation (\ref{scalarN}):
$$
 \iota=(1-p)n^2+\sum_{j=0}^p\sum_{\nu=0}^{n_j}{m_{j\nu}}^2\;.
$$
The differential equation is called {\it rigid} if it is uniquely determined by the Riemann scheme
(\ref{scalarRS}),
and the criterion of rigidity is given by the index of rigidity $\iota$ in a similar manner
as the case of Fuchsian systems.

The basic operations in the Katz theory are also defined for scalar equations.

{\bf i.}
The {\it addition} with parameter $(\alpha_1,\alpha_2,\dots,\alpha_p)\in\mathbb{C}^n$
is the operation which sends solutions $y(x)$ of (\ref{scalarN}) to
$$
 y(x)\prod_{j=1}^p(x-a_j)^{\alpha_j}\;,
$$
and is denoted by $add_{(\alpha_1,\alpha_2,\dots,\alpha_p)}$.
By this operation, the Riemann scheme (\ref{scalarRS}) changes to
$$
 \left\{
  \begin{matrix}
   x=a_1&x=a_2&\cdots&x=a_p&x=\infty\\
   [\lambda_{10}+\alpha_1]_{(m_{10})}
   &[\lambda_{20}+\alpha_2]_{(m_{20})}&\cdots
    &[\lambda_{p0}+\alpha_p]_{(m_{p0})}
    &[\lambda_{00}+\alpha_0]_{(m_{00})}\\
   [\lambda_{11}+\alpha_1]_{(m_{11})}
   &[\lambda_{21}+\alpha_2]_{(m_{21})}&\cdots
    &[\lambda_{p1}+\alpha_p]_{(m_{p1})}
    &[\lambda_{01}+\alpha_0]_{(m_{01})}\\
   \vdots&\vdots&&\vdots&\vdots\\
   [\lambda_{1n_1}+\alpha_1]_{(m_{1n_1})}
   &[\lambda_{2n_2}+\alpha_2]_{(m_{2n_2})}&\cdots
    &[\lambda_{pn_p}+\alpha_p]_{(m_{pn_p})}
    &[\lambda_{0n_0}+\alpha_0]_{(m_{0n_0})}
  \end{matrix}
 \right\},
$$
where we set
$$
 \alpha_0=-\sum_{j=1}^p\alpha_j\;.
$$
The addition does not change the spectral type.

{\bf ii.}
The {\it middle convolution} with parameter $\lambda$ is the operation which sends solutions
$y(x)$ of (\ref{scalarN}) to
$$
 z(x)=\int_{\Delta}y(t)(t-x)^{\lambda-1}\,dt
$$
and is denoted by $mc_{\lambda}$.
Then $z(x)$ satisfies another scalar Fuchsian differential equation,
which is said to be the result of the middle convolution.
In order to describe the Riemann scheme of the result of the middle convolution,
we use the following convention.
We set
$\lambda_{j0}=0$ for each $j=1,2,\dots,p$.
Then, if no exponent at $x=a_j$ is $0$, we have $m_{j0}=0$.
Also we set $\lambda_{00}=\lambda+1$, where $\lambda$ is the parameter of
the middle convolution,
and hence $m_{00}$ can take value $0$.
Then, by the middle convolution $mc_{\lambda}$,
the Riemann scheme changes as
\begin{equation}
\label{scalarMCRS}
 \begin{aligned}
  &
    \left\{
  \begin{matrix}
   x=a_1&x=a_2&\cdots&x=a_p&x=\infty\\
   [0]_{(m_{10})}&[0]_{(m_{20})}&\cdots
    &[0]_{(m_{p0})}&[\lambda+1]_{(m_{00})}\\
   [\lambda_{11}]_{(m_{11})}&[\lambda_{21}]_{(m_{21})}&\cdots
    &[\lambda_{p1}]_{(m_{p1})}&[\lambda_{01}]_{(m_{01})}\\
   \vdots&\vdots&&\vdots&\vdots\\
   [\lambda_{1n_1}]_{(m_{1n_1})}&[\lambda_{2n_2}]_{(m_{2n_2})}&\cdots
    &[\lambda_{pn_p}]_{(m_{pn_p})}&[\lambda_{0n_0}]_{(m_{0n_0})}
  \end{matrix}
 \right\}\\
 &\to
   \left\{
  \begin{matrix}
   x=a_1&x=a_2&\cdots&x=a_p&x=\infty\\
   [0]_{(m_{10}-d)}&[0]_{(m_{20}-d)}&\cdots
    &[0]_{(m_{p0}-d)}&[1-\lambda]_{(m_{00}-d)}\\
   [\lambda_{11}+\lambda]_{(m_{11})}&[\lambda_{21}+\lambda]_{(m_{21})}&\cdots
    &[\lambda_{p1}+\lambda]_{(m_{p1})}&[\lambda_{01}-\lambda]_{(m_{01})}\\
   \vdots&\vdots&&\vdots&\vdots\\
   [\lambda_{1n_1}+\lambda]_{(m_{1n_1})}&[\lambda_{2n_2}+\lambda]_{(m_{2n_2})}&\cdots
    &[\lambda_{pn_p}+\lambda]_{(m_{pn_p})}&[\lambda_{0n_0}-\lambda]_{(m_{0n_0})}
  \end{matrix}
 \right\},
 \end{aligned}
\end{equation}
where we set
$$
 d=\sum_{j=0}^pm_{j0}-(p-1)n\;.
$$
In particular, the order of the result of the middle convolution becomes $n-d$.
Also we can read the change of the spectral type by (\ref{scalarMCRS}).

It is also shown that an irreducible rigid scalar Fuchsian differential equation
can be obtained from a first order scalar Fuchsian differential equation
by a finite iteration of additions and middle convolutions.
The chain of additions and middle convolutions can be obtained in a similar manner
as in the case of Fuchsian systems.

\subsection{Application of Katz theory to CFT equations: solved cases}

In Section 3 we gave several scalar differential equations satisfied by some conformal blocks.
In this subsection, we explain that these differential equations together with their
integral representations of solutions can be obtained by applying the Katz theory.
Here we use the Katz theory for Fuchsian systems just for explanation.
The same results can be obtained by using the Katz theory for scalar Fuchsian equations.

\subsubsection{Second order Virasoro ODE}

In Section 3.1, we considered the second order differential equation
(\ref{rank2}) satisfied by (a gauge transform of) the conformal block $\mathcal{G}_{a_M}(x)$.
The Riemann scheme of the equation is given by
(\ref{Riem_not_2}).
We see the spectral type of the equation (\ref{rank2}) is $(11,11,11)$,
and then the index of rigidity is $\iota=(1-2)\times 2^2+3\times(1^2+1^2)=2$,
which implies the equation is rigid.
We have a chain
$$
 (1,1,1)\to(1^*1,1^*1,11)
$$
which connect the equation (\ref{rank2}) to rank one equation.

We shall obtain the equation (\ref{rank2})
by constructing a Fuchsian system of rank two by operating the middle convolution to
a equation of rank 1
\begin{equation}
\label{rank1S4.3.1}
 \frac{dy}{dx}=\left(\frac{\nu_1}{x}+\frac{\nu_2}{x-1}\right)y
\end{equation}
of spectral type $(1,1,1)$.

Let $\sigma_1$ be a new parameter.
We operate the middle convolution $mc_{\sigma_1}$ to (\ref{rank1S4.3.1}).
Then we get the Fuchsian system
\begin{equation}
\label{rank2S4.3.1}
 \frac{dY}{dx}=\left(\frac{A_1}{x}+\frac{A_2}{x-1}\right)Y
\end{equation}
of rank 2,
where
$$
 A_1=\begin{pmatrix}\nu_1+\sigma_1&\nu_2\\ 0&0\end{pmatrix},\
 A_2=\begin{pmatrix}0&0\\ \nu_1&\nu_2+\sigma_1\end{pmatrix}.
$$
We have
$$
 \begin{aligned}
  {\rm Ev}(A_1)&:0\;,\nu_1+\sigma_1\;,\\
  {\rm Ev}(A_2)&:0\;,\nu_2+\sigma_1\;,\\
  {\rm Ev}(A_0)&:-\sigma_1\;,-\nu_1-\nu_2-\sigma_1\;,
 \end{aligned}
$$
where ${\rm Ev}(A)$ denotes the eigenvalues.

An integral representation of solutions of (\ref{rank2S4.3.1}) is obtained
from the solution
$$
 y(x)=x^{\nu_1}(x-1)^{\nu_2}
$$
of (\ref{rank1S4.3.1}) by applying the Riemann-Liouville transform:
\begin{equation}
\label{integral4.3.1}
 \begin{aligned}
  Y(x)&=\int_{\Delta}t^{\nu_1}(t-1)^{\nu_2}(t-x)^{\sigma_1}\,\vec\eta\;,\\
  &\quad\vec\eta=\left(\frac{dt}{t},\frac{dt}{t-1}\right)^T\;.
 \end{aligned}
\end{equation}
In particular, the second entry $y_2(x)$ of $Y(x)$ is expressed as
\begin{equation}
\label{y2S4.3.1}
 y_2(x)=\int_{\Delta}t^{\nu_1}(t-1)^{\nu_2-1}(t-x)^{\sigma_1}\,dt\;.
\end{equation}
Comparing this with (\ref{GaussInt}),
we have a correspondence of parameters:
$$
 \nu_1=\beta-\gamma\;,\
 \nu_2=\gamma-\alpha\;,\
 \sigma_1=-\beta\;.
$$
By analyzing the integral (\ref{y2S4.3.1}),
we see that the local behaviors are given just by the Riemann scheme (\ref{Riem_not_2}).
Then, thanks to the rigidity,
we know that $y_2(x)$ satisfies the differential equation (\ref{rank2}),
and thus we get the integral representation (\ref{GaussInt})
of solutions of (\ref{rank2}).

\subsubsection{Third-order ODE for $\mathcal{B}^{(0,s)}_M(x)$}

The differential equation (\ref{Pochgamer}) satisfied by the gauge transform of
$\mathcal{B}^{(0,s)}_M(x)$
has spectral type $(111,21,111)$, which can be read from the Riemann scheme
(\ref{Riem_not_3}).
The index of rigidity is $\iota=(1-2)\times3^2+2(1^2+1^2+1^2)+(2^2+1^2)=2$,
and hence the differential equation is rigid.
Them we have a chain
$$
 (1,1,1)\xrightarrow{mc}(1^*1,1^*1,11)\xrightarrow{add}
 (11,1^*1,11)\xrightarrow{mc}(1^*11,2^*1,111)
$$
which connect (\ref{Pochgamer}) to rank one equation.
Since we have already obtained the Fuchsian system of spectral type $(1^*1,1^*1,11)$
in the above,
we may start from that system (\ref{rank2S4.3.1}).

We operate the addition $add_{(\nu_3,0)}$ in order to kill the kernel of $A_1$,
and then get $(A_1',A_2)$ with
$$
 A_1'=A_1+\nu_3=\begin{pmatrix}\nu_1+\sigma_1+\nu_3&\nu_2\\ 0&\nu_3\end{pmatrix}.
$$
Let $\sigma_2$ be a new parameter.
We operate the middle convolution $mc_{\sigma_2}$ to the system with residue matrices
$(A_1',A_2)$.
First we get the pair $(B_1,B_2)$ of $4\times4$ matrices:
$$
 B_1=\begin{pmatrix}A_1'+\sigma_2&A_2\\ O&O\end{pmatrix},\
 B_2=\begin{pmatrix}O&O\\ A_1'&A_2+\sigma_2\end{pmatrix},
$$
and then the Fuchsian system
\begin{equation}
\label{Y(x)}
 \frac{dY}{dx}=\left(\frac{B_1}{x}+\frac{B_2}{x-1}\right)Y
\end{equation}
of rank 4.
Since $A_2$ has the kernel $\langle(\nu_2+\sigma_2,-\nu_1)^T\rangle$ and
${\rm Ker}(B_1+B_2)=\{0\}$,
we have
$$
 \mathfrak{k}=\langle v\rangle,\
 \mathfrak{l}=\{0\}
$$
with
$$
 v=(0,0,\nu_2+\sigma_2,-\nu_1)^T.
$$
Then the middle convolution $mc_{\sigma_2}$ is realized by taking the actions of
$B_1,B_2$ on the quotient space
$\mathbb{C}^4/(\mathfrak{k}+\mathfrak{l})=\mathbb{C}^4/\langle v\rangle$.
To obtain the actions, we set
$P=(v,e_2,e_1,e_3)^T$,
where $(e_1,e_2,e_3,e_4)^T$ is the standard basis for $\mathbb{C}^4$.
Then we have
$$
 P^{-1}B_1P=
 \left(
 \begin{array}{c|ccc}
  0&*&*&*\\ \hline
  0&&&\\
  0&&C_1&\\
  0&&&
 \end{array}
 \right),\
 P^{-1}B_2P=
 \left(
 \begin{array}{c|ccc}
  \sigma_2&*&*&*\\ \hline
  0&&&\\
  0&&C_2&\\
  0&&&
 \end{array}
 \right)
$$
with
$$
 C_1=
  \begin{pmatrix}
   \nu_3+\sigma_2&0&\nu_1\\
   \nu_2&\nu_1+\nu_3+\sigma_1+\sigma_2&0\\
   0&0&0
  \end{pmatrix},\
 C_2=
  \begin{pmatrix}
   0&0&0\\
   0&0&0\\
   \frac{\nu_1\nu_2+\nu_2\nu_3+\sigma_1\nu_3}{\nu_1}
   &\nu_1+\nu_3+\sigma_1&\nu_2+\sigma_1+\sigma_2
  \end{pmatrix}.
$$
Thus, as a result of the middle convolutions, we get the
Fuchsian system
\begin{equation}
\label{FuchsSystem3}
 \frac{d}{dx}Z(x)= \left(\frac{C_1}{x}+\frac{C_2}{x-1}\right)Z(x)
\end{equation}
of rank 3.
The spectral type of (\ref{FuchsSystem3}) is $(1^*11,2^*1,111)$,
and, as is directly computed,
$$
 \begin{aligned}
  {\rm Ev}(C_1)&:0\;,\nu_3+\sigma_2\;,\nu_1+\nu_3+\sigma_1+\sigma_2\;,\\
  {\rm Ev}(C_2)&:0\;,0\;,\nu_2+\sigma_1+\sigma_2\;,\\
  {\rm Ev}(C_0)&:-\sigma_2\;,-\nu_3-\sigma_1-\sigma_2\;,-\nu_1-\nu_2-\nu_3-\sigma_1-\sigma_2\;.
 \end{aligned}
$$
Here we note that the unknown vector function $Z(x)=(z_1(x),z_2(x),z_3(x))^T$
in (\ref{FuchsSystem3})
is obtained by
$$
 PY(x)=\begin{pmatrix}*\\ \hline Z(x)\end{pmatrix}\;,
$$
where $Y(x)$ is the unknown vector function of the system (\ref{Y(x)}).
In particular, we have
\begin{equation}
\label{z1y2}
 z_1(x)=y_2(x)\;.
\end{equation}

We can derive from the system (\ref{FuchsSystem3})
a scalar differential equation,
which coincides with the generalized hypergeometric equation (\ref{Pochgamer}).
We consider the scalar differential equation satisfied by the first entry $z_1(x)$ of $Z(x)$.
By using the chain of transformations (\ref{gauge}) and (\ref{RiemannLiouville}),
we can obtain the integral representation of solutions $Z(x)$,
and hence of $z_1(x)$.
Namely we have
\begin{eqnarray}
 x^{\nu_1}(x-1)^{\nu_2} &\xrightarrow{mc_{\sigma_1}}
 & \int_{\delta} t_1^{\nu_1}(t_1-1)^{\nu_2}(t_1-x)^{\sigma_1}\,\vec\eta
 \xrightarrow{add_{(\nu_3,0)}}
 x^{\nu_3}\int_{\delta}t_1^{\nu_1}(t_1-1)^{\nu_2}(t_1-x)^{\sigma_1}\,\vec\eta\nonumber \\
 &\xrightarrow{mc_{\sigma_2}}
 &\int_{\Delta}t_1^{\nu_1}(t_1-1)^{\nu_2}(t_1-t_2)^{\sigma_1}t_2^{\nu_3}(t_2-x)^{\sigma_2}\,
 Q\vec\zeta=Z(x)\;,
\end{eqnarray}
where
\begin{equation}
\label{etazeta}
 \begin{aligned}
  \vec\eta&=\left(\frac{dt_1}{t_1},\frac{dt_1}{t_1-1}\right)^T,\\
  \vec\zeta&=\left(\frac{dt_1\wedge dt_2}{t_1t_2},\frac{dt_1\wedge dt_2}{(t_1-1)t_2},
  \frac{dt_1\wedge dt_2}{t_1(t_2-1)},\frac{dt_1\wedge dt_2}{(t_1-1)(t_2-1)}\right)^T,
 \end{aligned}
\end{equation}
and
$$
 Q=\begin{pmatrix}0&1&0&0\\ 0&0&1&0\\ 0&0&0&1\end{pmatrix}P\;.
$$
Then, noting (\ref{z1y2}) and (\ref{etazeta}),
we get the integral representation of $z_1(x)$:
\begin{equation}
\label{z1integral}
 z_1(x)
 =
 \int_{\Delta}t_1^{\nu_1}(t_1-1)^{\nu_2-1}(t_1-t_2)^{\sigma_1}
 t_2^{\nu_3-1}(t_2-x)^{\sigma_2}\,dt_1\wedge dt_2\;.
\end{equation}
By analyzing this integral, we can obtain the Riemann scheme of the scalar differential equation
for $z_1(x)$:
$$
 \left\{
  \begin{matrix}
   x=0&x=1&x=\infty\\
   0&0&-\sigma_2\\
   \nu_3+\sigma_2&1&-(\nu_3+\sigma_1+\sigma_2)\\
   \nu_1+\nu_3+\sigma_1+\sigma_2+1&\nu_2+\sigma_1+\sigma_2+1
   &-(\nu_1+\nu_2+\nu_3+\sigma_1+\sigma_2)
  \end{matrix}
 \right\}.
$$
Then we see that the substitution
\begin{equation}
\label{substitutionr3}
 \nu_1 =p_2 - q_2,\ 
 \nu_2 = q_2-p_3, \
 \nu_3 =p_1-q_1+1,\
 \sigma_1 =q_1-p_2-1,\
 \sigma_2 = -p_1
\end{equation}
establishes the correspondence of the generalized hypergeometric equation
(\ref{Pochgamer})
and the scalar equation for $z_1(x)$,
so that we have
$z_1(x)=G_{\vec{\alpha}_M}(x)$.

\subsubsection{Fourth-order ODE for $\mathcal{B}^{(1,-b)}_M(x)$}

The spectral type of the differential equation satisfied by
$\mathcal{B}^{(1,-b)}_M(x)$ is $(211,211,211)$,
which is readily read from the Riemann scheme (\ref{rank4RS}).
The index of rigidity is $\iota=(1-2)\times4^2+3(2^2+1^2+1^2)=2$,
and hence the differential equation is rigid.
Then we have a chain
$$
 (1,1,1)\xrightarrow{mc}(1^*1,1^*1,11)\xrightarrow{add}(11,11,11)
 \xrightarrow{mc}(2^*11,2^*11,211)
$$
which connect the differential equation to a rank one equation.
We can start from the system (\ref{rank2S4.3.1}).

We operate the addition $add_{(\nu_3,\nu_4)}$ to get $(A_1',A_2')$ with
$$
 \begin{aligned}
  A_1'&=A_1+\nu_3=\begin{pmatrix}\nu_1+\sigma_1+\nu_3&\nu_2\\ 0&\nu_3\end{pmatrix},\\
  A_2'&=A_2+\nu_4=\begin{pmatrix}\nu_4&0\\ \nu_1&\nu_2+\sigma_1+\nu_4\end{pmatrix}.
 \end{aligned}
$$
Then we have ${\rm Ker}A_1'={\rm Ker}A_2'=\{0\}$.
Let $\sigma_2$ be a new parameter.
We operate the middle convolution $mc_{\sigma_2}$ to the system with the residue matrices
$(A_1',A_2')$.
Then we get
$$
 B_1=\begin{pmatrix}A_1'+\sigma_2&A_2'\\ O&O\end{pmatrix},\
 B_2=\begin{pmatrix}O&O\\ A_1'&A_2'+\sigma_2\end{pmatrix},
$$
so that we see ${\rm Ker}(B_1+B_2)=\{0\}$.
Therefore $\mathfrak{k}=\mathfrak{l}=\{0\}$,
and hence the pair $(B_1,B_2)$ gives the middle convolution:
\begin{equation}
\label{211211211system}
 \frac{dZ}{dx}=\left(\frac{B_1}{x}+\frac{B_2}{x-1}\right)Z.
\end{equation}
The eigenvalues of $B_1,B_2$ and $B_0=-B_1-B_2$ are
\begin{equation}
\label{211211211RS}
 \begin{aligned}
  {\rm Ev}(B_1)&:0\;,0\;,\nu_3+\sigma_2\;,\nu_1+\nu_3+\sigma_1+\sigma_2\;,\\
  {\rm Ev}(B_2)&:0\;,0\;,\nu_4+\sigma_2\;,\nu_2+\nu_4+\sigma_1+\sigma_2\;,\\
  {\rm Ev}(B_0)&:-\sigma_2\;,-\sigma_2\;,-\nu_3-\nu_4-\sigma_1-\sigma_2\;,
  -\nu_1-\nu_2-\nu_3-\nu_4-\sigma_1-\sigma_2\;,
 \end{aligned}
\end{equation}
and hence the spectral type of the last system (\ref{211211211system}) is $(2^*11,2^*11,211)$. 

The system (\ref{211211211system}) can be transformed to a scalar differential equation of fourth order for each of the components $z_i(x)$, $i=1,\cdots,4$, of $Z(x)$.
Let us consider the case that we take $z_1(x)$ as the unknown of the scalar equation. We know a priori that the scalar equation is Fuchsian with singular points at $x=0,1,\infty$ and possibly at some apparent singular points,
and that the Riemann scheme is obtained from (\ref{211211211RS})
by translating some eigenvalues by integers.
Local analysis at each singular point gives the characteristic exponents.
We consider at $x=1$.
Put
$$
 Z(x)=(x-1)^{\rho}\sum_{n=0}^{\infty}Z_n(x-1)^n
$$
into the system (\ref{211211211system}).
Then we see that $\rho$ is an eigenvalue of $B_2$ and $Z_0$ is an eigenvector
for $\rho$.
Moreover, we have
$$
 \begin{aligned}
  Z_1&=(B_2-(\rho+1))^{-1}B_1(-Z_0)\;,\\
  Z_2&=(B_2-(\rho+2))^{-1}B_1(-Z_1+Z_0)\;.
 \end{aligned}
$$
For the double eigenvalue $\rho=0$,
we see that there is an eigenvector of the form $(*,*,*,*)^{{\rm T}}$,
where $*$ stands for a non-zero entry.
This implies that we have characteristic exponents $0$ and $1$ at $x=1$.
For the eigenvalue $\rho=\nu_4+\sigma_2$,
we have
$$
 Z_0=(0,0,*,*)^{{\rm T}},\
 Z_1=(*,*,*,*)^{{\rm T}}.
$$
Then, for the scalar equation satisfied by $z_1$,
the corresponding characteristic exponent becomes $\nu_4+\sigma_2+1$,
because the first entry of $Z_0$ is zero while one of $Z_1$ is non-zero.
For the eigenvalue $\rho=\nu_2+\nu_4+\sigma_1+\sigma_2$,
we have
$$
 Z_0=(0,0,0,*)^{{\rm T}},\
 Z_1=(0,*,*,*)^{{\rm T}},\
 Z_2=(*,*,*,*)^{{\rm T}}.
$$
Then, in a similar reason, the corresponding characteristic exponent becomes
$\nu_2+\nu_4+\sigma_1+\sigma_2+2$.
Similarly we can calculate the characteristic exponents at $x=0$ and $x=\infty$.
Thus we obtain a temporary Riemann scheme
\begin{equation}
\label{211211211RSz4}
 \left\{
  \begin{matrix}
   x=0&x=1&x=\infty\\
   0&0&-\sigma_2\\
   1&1&-\sigma_2+1\\
   \nu_3+\sigma_2&\nu_4+\sigma_2+1&-\nu_3-\nu_4-\sigma_1-\sigma_2\\
   \nu_1+\nu_3+\sigma_1+\sigma_2&\nu_2+\nu_4+\sigma_1+\sigma_2+2
   &-\nu_1-\nu_2-\nu_3-\nu_4-\sigma_1-\sigma_2
  \end{matrix}
 \right\}.
\end{equation}
It can be shown that the Fuchs relation
(\ref{FuchsRelation}) holds for (\ref{211211211RSz4}),
which implies that there is no apparent singular point and that (\ref{211211211RSz4})
is the actual Riemann scheme for the scalar differential equation satisfied by $z_1(x)$.

Now we compute the integral representation of solutions of the system.
We can start from the integral representation
(\ref{integral4.3.1}) for the system of spectral type $(1^*1,1^*1,11)$.
By the addition $add_{(\nu_3,\nu_4)}$ we get
$$
 x^{\nu_3}(x-1)^{\nu_4}\int_{\Delta}t_1^{\nu_1}(t_1-1)^{\nu_2}(t_1-x)^{\sigma_1}\,\vec\eta
$$
with
$$
 \vec\eta=\left(\frac{dt_1}{t_1},\frac{dt_1}{t_1-1}\right)^T.
$$
Then by the middle convolution $mc_{\sigma_2}$, we obtain
the integral representation
$$
 Z(x)=\int_{\Delta}t_1^{\nu_1}(t_1-1)^{\nu_2}(t_1-t_2)^{\sigma_1}
 t_2^{\nu_3}(t_2-1)^{\nu_4}(t_2-x)^{\sigma_2}\,\vec\zeta
$$
with
$$
 \begin{aligned}
  \vec\zeta&=
  \left(\vec\eta\wedge\frac{dt_2}{t_2},\vec\eta\wedge\frac{dt_2}{t_2-1}\right)^{{\rm T}}\\
  &=\left(\frac{dt_1\wedge dt_2}{t_1t_2},\frac{dt_1\wedge dt_2}{(t_1-1)t_2},
  \frac{dt_1\wedge dt_2}{t_1(t_2-1)},\frac{dt_1\wedge dt_2}{(t_1-1)(t_2-1)}\right)^{{\rm T}}.
 \end{aligned}
$$
In particular, we can derive the integral representation of $z_1(x)$:
\begin{equation}
\label{z_1Integral}
 z_1(x)=\int_{\Delta}t_1^{\nu_1-1}(t_1-1)^{\nu_2}(t_1-t_2)^{\sigma_1}
 t_2^{\nu_3-1}(t_2-1)^{\nu_4}(t_2-x)^{\sigma_2}\,dt_1\wedge dt_2\;.
\end{equation}

Now we shall relate the above results to ones in \cite{BEFS16}.
In order to relate the system (\ref{211211211system}) to the differential equation (4.21)
in \cite{BEFS16},
we operate the addition
\begin{equation}
\label{adB}
 (B_1,B_2)\to(B_1+\alpha,B_2+\beta)\;.
\end{equation}
Then the Riemann scheme for the scalar equation of $z_1$ becomes
$$
 \left\{
  \begin{matrix}
   x=0&x=1&x=\infty\\
   \alpha&\beta&-\sigma_2-\alpha-\beta\\
   \alpha+1&\beta+1&-\sigma_2-\alpha-\beta+1\\
   \nu_3+\sigma_2+\alpha&\nu_4+\sigma_2+\beta+1&-\nu_{34}-\sigma_{12}-\alpha-\beta\\
   \nu_{13}+\sigma_{12}+\alpha&\nu_{24}+\sigma_{12}+\beta+2
   &-\nu_{1234}-\sigma_{12}-\alpha-\beta
  \end{matrix}
 \right\},
$$
where we used the notation $\nu_{ij\cdots k}=\nu_i+\nu_j+\cdots+\nu_k,\sigma_{12}=\sigma_1+\sigma_2$.
We impose the condition
$$
 \left\{
  \begin{aligned}
   &-\sigma_2-\alpha-\beta=\gamma_3\;,\\
   &\nu_3+\sigma_2+\alpha=\alpha_1\;,\\
   &\nu_{13}+\sigma_{12}+\alpha=\alpha_2\;,\\
   &\nu_4+\sigma_2+\beta+1=\beta_1\;,\\
   &\nu_{24}+\sigma_{12}+\beta+2=\beta_2\;,\\
   &-\nu_{34}-\sigma_{12}-\alpha-\beta=\gamma_2\;,
  \end{aligned}
 \right.
$$
Then we have a unique solution
$$
 \left\{
  \begin{aligned}
   \nu_1&=-a_{R_2}b+1\;,\\
   \nu_3&=-a_{R_1}b+1\;,\\
   \nu_2&=\nu_4=\sigma_1=\sigma_2=b^2\;.
  \end{aligned}
 \right.
$$
Put this solution into the integral representation (\ref{z_1Integral})
and operate the gauge transform coming from the addition (\ref{adB}).
Then we get the integral representation
$$
 x^{\alpha}(x-1)^{\beta}
 \int_{\Delta}
 {t_1}^{-a_{R_2}b}(t_1-1)^{b^2}(t_1-t_2)^{b^2}{t_2}^{-a_{R_1}b}(t_2-1)^{b^2}(t_2-x)^{b^2}
 dt_1\wedge dt_2\;,
$$
which turns out to coincide with the integral (5.2) in \cite{BEFS16}.
By using this integral, we can compute the monodromy and the monodromy invariant
Hermitian form.

The system (\ref{211211211system}) is essentially the Okubo system which is obtained in 
Theorem II in \cite{H1} with $m=2$,
and the monodromy representation and the monodromy invariant Hermitian form are also
obtained in \cite{H2}.
There is some difference in the normalization, however,
we arrive at almost same conclusion.

\section{Application of Katz theory: the $\cB_{M}^{(1,s)}(z)$ case}
\label{applications2}

We now apply the Katz theory to new cases. 
In \cite{BCES17} we obtained a sixth-order differential equation for the conformal block  $\cB_{M}^{(1,s)}(x)$ that we were not able to solve.

\subsection{Chain of addition and middle convolution transformations}

As we have argued in Section \ref{dimorigid} and proven in \cite{BCES17}, the Fuchsian system
satisfied by $\cB_{M}^{(1,s)}(x)$ is of spectral type $(222,321,222)$,
which has the index of rigidity $\iota=(1-2)\times6^2+2(2^2+2^2+2^2)+(3^2+2^2+1^2)=2$
and hence is rigid.
This system is obtained by the following chain
\begin{equation}
\label{rank6chain1}
 \begin{aligned}
  &(1,1,1)\xrightarrow{mc}(1^*1,1^*1,11)\xrightarrow{add}(11,11,11)
  \xrightarrow{mc}(2^*11,2^*11,211)\\
  &\xrightarrow{add}(21^*1,21^*1,21^*1)=(1^*21,1^*21,1^*21)
  \xrightarrow{mc}(2^*21,2^*21,221)\\
  &\xrightarrow{add}(221^*,2^*21,221^*)=(1^*22,2^*21,1^*22)
  \xrightarrow{mc}(2^*22,3^*21,222)\;.
 \end{aligned}
\end{equation}
The system of spectral type $(2^*11,2^*11,211)$
coincides with the system (\ref{211211211system}).
Then we can obtain the Fuchsian system of spectral type $(222,321,222)$
by starting from (\ref{211211211system})
and operating two additions and two middle convolutions.
However,
in each of the two middle convolutions,
we should take a linear transform to get the action on the quotient space,
which involves non-canonical basis.
Then usually the explicit forms of the system and the vector of the twisted cocycles
in the integral representation become complicated.
In order to avoid such inconveniences,
we apply the Katz theory for scalar differential equations,
which is explained in Section 4.2.

The chain (\ref{rank6chain1}) is the same for scalar case.
We give the Riemann schemes and the integral representations in each step.

\noindent
\underline{Spectral type $(1,1,1)$}:
$$
 \left\{
  \begin{matrix}x=0&x=1&x=\infty\\ \nu_1&\nu_2&-\nu_{12}\end{matrix}
 \right\},
$$
$$
 x^{\nu_1}(x-1)^{\nu_2}.
$$
$\xrightarrow{mc_{\sigma_1}}$

\noindent
\underline{Spectral type $(1^*1,1^*1,11)$}:
$$
 \left\{
  \begin{matrix}
   x=0&x=1&x=\infty\\
   0&0&1-\sigma_1\\
   \nu_1+\sigma_1&\nu_2+\sigma_1&-\nu_{12}-\sigma_1
  \end{matrix}
 \right\},
$$
$$
 \int_{\Delta}t_1^{\nu_1}(t_1-1)^{\nu_2}(t_1-t_2)^{\sigma_1-1}\,dt_1\;.
$$
$\xrightarrow{add_{(\nu_3,\nu_4)}}$

\noindent
\underline{Spectral type $(11,11,11)$}:
$$
 \left\{
  \begin{matrix}
   x=0&x=1&x=\infty\\
   \nu_3&\nu_4&1-\nu_{34}-\sigma_1\\
   \nu_{13}+\sigma_1&\nu_{24}+\sigma_1&-\nu_{1234}-\sigma_1
  \end{matrix}
 \right\},
$$
$$
 x^{\nu_3}(x-1)^{\nu_4}\int_{\Delta}t_1^{\nu_1}(t_1-1)^{\nu_2}(t_1-x)^{\sigma_1-1}\,dt_1\;.
$$
$\xrightarrow{mc_{\sigma_2}}$

\noindent
\underline{Spectral type $(2^*11,2^*11,211)$}:
$$
 \left\{
  \begin{matrix}
   x=0&x=1&x=\infty\\
   0&0&1-\sigma_2\\
   1&1&2-\sigma_2\\
   \nu_3+\sigma_2&\nu_4+\sigma_2&1-\nu_{34}-\sigma_{12}\\
   \nu_{13}+\sigma_{12}&\nu_{24}+\sigma_{12}&-\nu_{1234}-\sigma_{12}
  \end{matrix}
 \right\},
$$
$$
 \int_{\Delta}t_1^{\nu_1}(t_1-1)^{\nu_2}(t_1-t_2)^{\sigma_1-1}
 t_2^{\nu_3}(t_2-1)^{\nu_4}(t_2-x)^{\sigma_2-1}\,dt_1\wedge dt_2\;.
$$
$\xrightarrow{add_{(-\nu_3-\sigma_2,-\nu_4-\sigma_2)}}$

\noindent
\underline{Spectral type $(1^*21,1^*21,21^*1)$}:
$$
 \left\{
  \begin{matrix}
   x=0&x=1&x=\infty\\
   0&0&\nu_{34}+\sigma_2+1\\
   -\nu_3-\sigma_2&-\nu_4-\sigma_2&\nu_{34}+\sigma_2+2\\
   -\nu_3-\sigma_2+1&-\nu_4-\sigma_2+1&\sigma_2-\sigma_1+1\\
   \nu_1+\sigma_1&\nu_2+\sigma_1&-\nu_{12}+\sigma_2-\sigma_1
  \end{matrix}
 \right\},
$$
$$
 x^{-\nu_3-\sigma_2}(x-1)^{-\nu_4-\sigma_2}
 \int_{\Delta}t_1^{\nu_1}(t_1-1)^{\nu_2}(t_1-t_2)^{\sigma_1-1}
 t_2^{\nu_3}(t_2-1)^{\nu_4}(t_2-x)^{\sigma_2-1}\,dt_1\wedge dt_2\;.
$$
$\xrightarrow{mc_{\sigma_2-\sigma_1}}$

\noindent
\underline{Spectral type $(2^*21,2^*21,221)$}:
$$
 \left\{
  \begin{matrix}
   x=0&x=1&x=\infty\\
   0&0&\sigma_1-\sigma_2+1\\
   1&1&\sigma_1-\sigma_2+2\\
   -\nu_3-\sigma_1&-\nu_4-\sigma_1&\nu_{34}+\sigma_1+1\\
   -\nu_3-\sigma_1+1&-\nu_4-\sigma_1+1&\nu_{34}+\sigma_1+2\\
   \nu_1+\sigma_2&\nu_2+\sigma_2&-\nu_{12}
  \end{matrix}
 \right\},
$$
$$
 \int_{\Delta}t_1^{\nu_1}(t_1-1)^{\nu_2}(t_1-t_2)^{\sigma_1-1}
 t_2^{\nu_3}(t_2-1)^{\nu_4}(t_2-t_3)^{\sigma_2-1}
 t_3^{-\nu_3-\sigma_2}(t_3-1)^{-\nu_4-\sigma_2}(t_3-x)^{\sigma_2-\sigma_1}
 \,dt_1\wedge dt_2\wedge dt_3\;.
$$
$\xrightarrow{add_{(-\nu_1-\sigma_2,0)}}$

\noindent
\underline{Spectral type $(221^*,2^*21,221^*)$}:
$$
 \left\{
  \begin{matrix}
   x=0&x=1&x=\infty\\
   0&0&\nu_1+\sigma_1+1\\
   -\nu_1-\sigma_2&1&\nu_1+\sigma_1+2\\
   -\nu_1-\sigma_2+1&-\nu_4-\sigma_1&\nu_{134}+\sigma_{12}+1\\
   -\nu_{13}-\sigma_{12}&-\nu_4-\sigma_1+1&\nu_{134}+\sigma_{12}+2\\
   -\nu_{13}-\sigma_{12}+1&\nu_2+\sigma_2&-\nu_2+\sigma_2
  \end{matrix}
 \right\},
$$
$$
 \begin{aligned}
  x^{-\nu_1-\sigma_2}
  &\int_{\Delta}t_1^{\nu_1}(t_1-1)^{\nu_2}(t_1-t_2)^{\sigma_1-1}
  t_2^{\nu_3}(t_2-1)^{\nu_4}(t_2-t_3)^{\sigma_2-1}\\
  &\quad\times
  t_3^{-\nu_3-\sigma_2}(t_3-1)^{-\nu_4-\sigma_2}(t_3-x)^{\sigma_2-\sigma_1}
  \,dt_1\wedge dt_2\wedge dt_3\;.
 \end{aligned}
$$
$\xrightarrow{mc_{-\nu_2+\sigma_2-1}}$

\noindent
\underline{Spectral type $(2^*22,3^*21,222)$}:
$$
 \left\{
  \begin{matrix}
   x=0&x=1&x=\infty\\
   0&0&\nu_2-\sigma_2+2\\
   1&1&\nu_2-\sigma_2+3\\
   -\nu_{12}-1&2&\nu_{12}+\sigma_1-\sigma_2+2\\
   -\nu_{12}&-\nu_{24}-\sigma_1+\sigma_2-1&\nu_{12}+\sigma_1-\sigma_2+3\\
   -\nu_{123}-\sigma_1-1&-\nu_{24}-\sigma_1+\sigma_2&\nu_{1234}+\sigma_1+2\\
   -\nu_{123}-\sigma_1&2\sigma_2-1&\nu_{1234}+\sigma_1+3
  \end{matrix}
 \right\},
$$
$$
 \begin{aligned}
  &\int_{\Delta}t_1^{\nu_1}(t_1-1)^{\nu_2}(t_1-t_2)^{\sigma_1-1}
  t_2^{\nu_3}(t_2-1)^{\nu_4}(t_2-t_3)^{\sigma_2-1}
  t_3^{-\nu_3-\sigma_2}(t_3-1)^{-\nu_4-\sigma_2}\\
  &\quad\times
  (t_3-t_4)^{\sigma_2-\sigma_1-1}
  t_4^{-\nu_1-\sigma_2}(t_4-x)^{-\nu_2+\sigma_2-2}
  \,dt_1\wedge dt_2\wedge dt_3\wedge dt_4\;.
 \end{aligned}
$$
Thus we obtain an integral representation of solutions of the rigid scalar equation
of spectral type $(222,321,222)$.
We slightly change the parameters.
We use $\nu_2+1$ as a new $\nu_2$ and $\sigma_2-1$ as a new $\sigma_2$.
In these new parameters,
we have an integral representation of solutions
\begin{equation}
\label{rank6Integral}
 \begin{aligned}
  &\int_{\Delta}t_1^{\nu_1}(t_1-1)^{\nu_2-1}(t_1-t_2)^{\sigma_1-1}
  t_2^{\nu_3}(t_2-1)^{\nu_4}(t_2-t_3)^{\sigma_2}
  t_3^{-\nu_3-\sigma_2-1}(t_3-1)^{-\nu_4-\sigma_2-1}\\
  &\quad\times
  (t_3-t_4)^{\sigma_2-\sigma_1}
  t_4^{-\nu_1-\sigma_2-1}(t_4-x)^{-\nu_2+\sigma_2}
  \,dt_1\wedge dt_2\wedge dt_3\wedge dt_4\;.
 \end{aligned}
\end{equation}
of the scalar differential equation specified by the Riemann scheme
\begin{equation}
\label{rank6RS}
 \left\{
  \begin{matrix}
   x=0&x=1&x=\infty\\
   0&0&\nu_2-\sigma_2\\
   1&1&\nu_2-\sigma_2+1\\
   -\nu_{12}&2&\nu_{12}+\sigma_1-\sigma_2\\
   -\nu_{12}+1&-\nu_{24}-\sigma_1+\sigma_2+1&\nu_{12}+\sigma_1-\sigma_2+1\\
   -\nu_{123}-\sigma_1&-\nu_{24}-\sigma_1+\sigma_2+2&\nu_{1234}+\sigma_1+1\\
   -\nu_{123}-\sigma_1+1&2\sigma_2+1&\nu_{1234}+\sigma_1+2
  \end{matrix}
 \right\}.
\end{equation}
We recall that the conformal block $B^{(1,s)}_M(x)$ is associated to the Riemann scheme (\ref{Riem_rank6}). By sending the solution $B^{(1,s)}_M(x)$ to $x^{\alpha_1}(1-x)^{\beta_1}B^{(1,s)}_M(x)$ we can identify the corresponding Riemann scheme with the one above, provided that:
\begin{align}
\label{alphas_katz}
&\alpha_2-\alpha_1 = -\nu_{12}\;, \quad \alpha_3-\alpha_1=-\nu_{123}-\sigma_1 \;,\nonumber \\
&\beta_3-\beta_1 = -\nu_{24}-\sigma_1+\sigma_2+1\;, 
\quad \beta_2 - \beta_1 = 2 \sigma_2 + 1\;, \nonumber \\
&\gamma_1 + \alpha_1 + \beta_1 = 
 \nu_2 - \sigma_2\;, \quad  \gamma_2 + \alpha_1 + \beta_1 = \nu_{1234}+\sigma_1+1\;.
\end{align}

Using the (\ref{Riem_param_6}), we can relate the parameters
$(\nu_1,\nu_2,\sigma_1,\nu_3,\nu_4,\sigma_2)$, entering naturally in the Katz procedure,  
to the Toda parameters $(a_{R_1},a_{R_2},a_{L_1},a_{L_2},s,b)$, see Section \ref{cb_sem_deg}.  
\begin{align}
\nonumber 
 \nu_1 &= \frac{b}{3}  \left(-a_{L_1}+a_{L_2}+a_{R_1}-a_{R_2}+b-s\right)\;,\\
 \nu_2&= -1+\frac{b}{3} \left(a_{L_1} -a_{L_2} +2 a_{R_1} +a_{R_2}b-4 b+ s\right)\;,\nonumber \\
\sigma_1&= 1+ \frac{b}{3}  (-2 a_{L_1}-a_{L_2}-a_{R_1}+a_{R_2}+2 b+s)\;,\nonumber \\
 \nu_3&= -2+\frac{b}{3} \left(2 a_{L_1} +a_{L_2} +a_{R_1} +2 a_{R_2} -5 b- s\right)\;,\nonumber \\
 \nu_4&=-1+\frac{b}{3}  (a_{L_1}+2a_{L_2}-a_{R_1}-2 a_{R_2}-4 b+s)\;,
 \quad
 \sigma_2= b^2\;.
\label{rank6coeff}
\end{align}
The identification (\ref{alphas_katz}) is not unique. 
There are 12 possible identifications as we can arbitrarily associate local exponents with the same multiplicity. For instance, we could have replaced the two last equations of (\ref{alphas_katz}) by:
\begin{equation}
\gamma_1 + \alpha_1 + \beta_1 = 
 \nu_{1234}+\sigma_1+1\;, \quad  \gamma_2 + \alpha_1 + \beta_1 = \nu_2 - \sigma_2\;.
\label{alphas_katz_bis}
\end{equation}
The crucial observation in that replacing the last two equations in (\ref{alphas_katz}) by the ones in (\ref{alphas_katz_bis}) is equivalent to the following transformations of the Toda parameters:
\begin{equation}
(a_{R_{1}},a_{R_2})\to (a_{R_{1}},a_{R_2})\;, \quad (a_{L_{1}},a_{L_2})\to (a_{L_{1}}+a_{L_2}-Q\;, 2Q-a_{L_2})\;,
\end{equation}
that coincides with one of the Weyl transformations (\ref{weyl}) on the charge $\vec{\alpha}_L$. 
In general the possible $12$ transformations correspond to Weyl transformations of the charges $\vec{\alpha}_R$ and $\vec{\alpha}_L$. From the point of view of the Toda field theory, this means that all the possible identifications are equivalent.

\subsection{Cycles corresponding to local fundamental systems of solutions}
\label{cycles}

We study the local behavior of the integral (\ref{rank6Integral}).
We denote the integrand of (\ref{rank6Integral}) by $u$, and set
$$
 S=\bigcup_{j=1}^4\{t_j=0\}\cup\bigcup_{j=1}^{3}
 \{t_j=1\}\cup\bigcup_{j=1}^3\{t_j=t_{j+1}\}
 \cup\{t_4=x\}\subset\mathbb{C}^4\;,
$$
which is the set of branch points of $u$.
Set
$$
 X=\mathbb{C}^4\setminus S
$$
and let $\mathcal{L}$ be the local system on $X$ defined by the multi-valued function $u$.
In the integral (\ref{rank6Integral}),
we take $\Delta$ in the twisted homology group $H^{\ell f}_4(X,\mathcal{L})$
of locally finite cycles in $X$.
Now, we find that many resonances  occur among the exponents of the integrand of (\ref{rank6Integral}).
For example, the sum of the exponents of $t_2,t_2-t_3$ and $t_3$ is
$$
 \nu_3+\sigma_2+(-\nu_3-\sigma_2-1)=-1\;,
$$
which is an integer.
Then we call that the divisor $t_2=t_3=0$ or the sum of the exponents resonant.
In this way, we find the following divisors are resonant:
$$
 \begin{aligned}
  &t_2=t_3=0\;,\
  t_2=t_3=1\;,\
  t_2=t_3=\infty\;,\\
  &t_1=t_2=t_3=t_4=0\;,\
  t_1=t_2=t_3=t_4=\infty\;.
 \end{aligned}
$$
We have the inclusion map
$$
 i:H_4(X,\mathcal{L})\to H_4^{\ell f}(X,\mathcal{L})
$$
and the image of $i$ is said to be regularizable cycles (cf. \cite{kita}, \cite{HM}, \cite{MY}).
A cycle $\Delta$ is regularizable if and only if any divisor in the boundary of $\Delta$
is non-resonant.
We are going to study the asymptotic behaviors of the integral (\ref{rank6Integral})
with regularizable $\Delta$.

First we assume $0<x<1$.
Singular asymptotic behaviors at $x=0$ are obtained by the cycles contained in $0<t_4<x$.
The cycle $0<t_1<t_2<t_3<t_4<x$ has a boundary $t_1=t_2=t_3=t_4=0$, which is resonant.
Then it is not regularizable.
Cycles contained in $0<t_2<t_3<t_4<x$ with $t_1\not\in(0,t_2)$ are not regularizable, because they have the
resonant boundary $t_2=t_3=0$.
We consider cycles contained in $0<t_3<t_4<x$ with $t_2\not\in(0,t_3)$.
We change the variables of integration $t_3,t_4$ to $s_3,s_4$, respectively, where
$$
 t_3=xs_3\;,\
 t_4=xs_4\;.
$$
Then we have
$$
 \begin{aligned}
  &(t_2-t_3)^{\sigma_2}{t_3}^{-\nu_3-\sigma_2-1}(t_3-1)^{-\nu_4-\sigma_2-1}
  (t_3-t_4)^{-\sigma_1+\sigma_2}{t_4}^{-\nu_1-\sigma_2-1}(t_4-x)^{-\nu_2+\sigma_2}\,
  dt_3\wedge dt_4\\
  &=
  x^{-\nu_1-\nu_2-\sigma_1-\nu_3}(t_2-xs_3)^{\sigma_2}(1-xs_3)^{-\nu_4-\sigma_2-1}
  g(s_3,s_4)\,ds_3\wedge ds_4\;,
 \end{aligned}
$$
with some function $g$ in $s_3,s_4$.
Note that we obtain $0<s_3<s_4<1$ from $0<t_3<t_4<x$.
If the factor $(t_2-xs_3)^{\sigma_2}$ is expanded in Taylor series at $x=0$,
we have an asymptotic behavior at $x=0$ with exponent $-\nu_1-\nu_2-\nu_3-\sigma_1$.
Thus we should take $t_2>1$.
Then, for any choice of the interval to which $t_1$ belongs,
the cycle becomes regularizable.
There are four intervals $(-\infty,0),(0,1),(1,t_2),(t_2,\infty)$ for $t_1$,
and we have two independent relations of these intervals which are regarded as
twisted 1-chains.
Thus we obtain two independent regularizable cycles in $H^{\ell f}_4(X,\mathcal{L})$.
This implies that the integral (\ref{rank6Integral}) with regularizable $\Delta$
gives the asymptotic behavior with exponent $-\nu_1-\nu_2-\nu_3-\sigma_1$ at $x=0$,
and the dimension of the integrals of this behavior is two.
In terms of the Riemann scheme,
we have the characteristic exponents $-\nu_1-\nu_2-\nu_3-\sigma_1$
and $-\nu_1-\nu_2-\nu_3-\sigma_1+1$
at $x=0$.

In this way, we can specialize all cycles which represent the exponents in the Riemann scheme
(\ref{rank6RS}).

The exponents in the Riemann scheme (\ref{rank6RS}) of the Fuchsian differential equation
of spectral type $(222,321,222)$ are realized by the integral (\ref{rank6Integral})
with the following regularizable cycles $\Delta$.

We assume $0<x<1$. The exponents $0,1$ at $x=0$ are realized by linear combinations of the cycles
\begin{equation}
\label{rank6sol12} 
 \begin{aligned}
   \Delta_{01a}&=\{t_2<0<t_1<1<t_3<t_4\}\;,\\
   \Delta_{01b}&=\{0<t_2<t_1<1<t_3<t_4\}\;.
 \end{aligned}
\end{equation}
The exponents $-\nu_1-\nu_2,-\nu_1-\nu_2+1$ at $x=0$ 
are realized by linear combinations of the cycles
\begin{equation}
\label{rank6sol34} 
 \begin{aligned}
  \Delta_{02a}&=\{t_1<t_2<0,\ t_3>1,\ 0<t_4<x\}\;,\\
  \Delta_{02b}&=\{0<t_2<t_1<1<t_3,\ 0<t_4<x\}\;.
 \end{aligned}
\end{equation}

The exponents $-\nu_1-\nu_2-\nu_3-\sigma_1,-\nu_1-\nu_2-\nu_3-\sigma_1+1$ at $x=0$
are realized by linear combinations of the cycles
\begin{equation}
\label{rank6sol56}
 \begin{aligned}
  \Delta_{03a}&=\{t_2>1,\ 0<t_3<t_4<x,\ t_1<0\}\;,\\
  \Delta_{03b}&=\{t_2>1,\ 0<t_3<t_4<x,\ 0<t_1<1\}\;,\\
  \Delta_{03c}&=\{t_2>1,\ 0<t_3<t_4<x,\ 1<t_1<t_2\}\;,\\
  \Delta_{03d}&=\{t_2>1,\ 0<t_3<t_4<x,\ t_2<t_1\}\;, 
 \end{aligned}
\end{equation}
where two linear relations among these four cycles hold by Cauchy's theorem
on $t_1$-plane.

The exponents $0,1,2$ at $x=1$ are realized by linear combinations of the cycles
\begin{equation}
 \begin{aligned}
  \Delta_{11a}&=\{t_3<t_4<0<t_2<1<t_1\}\;,\\
  \Delta_{11b}&=\{t_4<0<t_3<1<t_1<t_2\}\;,\\
  \Delta_{11c}&=\{t_4<0<t_2<t_1<1<t_3\}\;,\\
  \Delta_{11d}&=\{t_4<t_3<0<t_1<1<t_2\}\;.
 \end{aligned}
 \label{rank6sol1}
\end{equation}
Remark
 that there are four cycles for holomorphic solutions at $x=1$, while
the dimension of the space of holomorphic solutions at $x=1$ is three.
Then these four cycles should be linearly dependent.
Later in Corollary we shall show the linear dependence by solving the connection problem.
The exponents $-\nu_2-\nu_4-\sigma_1+\sigma_2+1,-\nu_2-\nu_4-\sigma_1+\sigma_2+2$
at $x=1$ are realized by linear combinations of the cycles
\begin{equation}
\label{rank6sol12abcd}
 \begin{aligned}
  \Delta_{12a}&=\{t_2<0,\ x<t_4<t_3<1,\ t_1<t_2\}\;,\\
  \Delta_{12b}&=\{t_2<0,\ x<t_4<t_3<1,\ t_2<t_1<0\}\;,\\
  \Delta_{12c}&=\{t_2<0,\ x<t_4<t_3<1,\ 0<t_1<1\}\;,\\
  \Delta_{12d}&=\{t_2<0,\ x<t_4<t_3<1,\ 1<t_1\}\;,\\
 \end{aligned}
\end{equation}
where two linear relations among these four cycles hold by Cauchy's theorem
on $t_1$-plane.
The exponent $2\sigma_2+1$ at $x=1$ is realized by the cycle
\begin{equation}
\label{rank6sol13}
 \Delta_{13}=\{x<t_4<t_3<t_2<t_1<1\}\;.
\end{equation}

Next we assume $1<x$.

The exponents $\nu_2-\sigma_2,\nu_2-\sigma_2+1$ at $x=\infty$ are realized by linear combinations of the cycles
\begin{equation}
\label{rank6soli1}
 \begin{aligned}
  \Delta_{\infty1a}&=\{t_2<0<t_4<t_3<1<t_1\}\;,\\
  \Delta_{\infty1b}&=\{0<t_4<t_3<1<t_1<t_2\}\;.
 \end{aligned}
\end{equation}
The exponents 
$\nu_1+\nu_2+\nu_3+\nu_4+\sigma_1+1,\nu_1+\nu_2+\nu_3+\nu_4+\sigma_1+2$
at $x=\infty$ are realized by linear combinations of the cycles
\begin{equation}
\label{rank6soli2}
 \begin{aligned}
  \Delta_{\infty2a}&=\{0<t_2<1,\ x<t_4<t_3,\ t_1<0\}\;,\\
  \Delta_{\infty2b}&=\{0<t_2<1,\ x<t_4<t_3,\ 0<t_1<t_2\}\;,\\
  \Delta_{\infty2c}&=\{0<t_2<1,\ x<t_4<t_3,\ t_2<t_1<1\}\;,\\
  \Delta_{\infty2d}&=\{0<t_2<1,\ x<t_4<t_3,\ 1<t_1\}\;,\\
 \end{aligned}
\end{equation}
where two linear relations among these four cycles hold by Cauchy's theorem
on $t_1$-plane.
The exponents $\nu_1+\nu_2+\sigma_1-\sigma_2,\nu_1+\nu_2+\sigma_1-\sigma_2+1$
at $x=\infty$ are realized by linear combinations of the cycles
\begin{equation}
\label{rank6soli3}
 \begin{aligned}
  \Delta_{\infty3a}&=\{t_2<t_1<0<t_3<1,\ t_4>x\}\;,\\
  \Delta_{\infty3b}&=\{0<t_3<1<t_1<t_2,\ t_4>x\}\;.
 \end{aligned}
\end{equation}

\subsection{Linear relations between twisted cycles}

We obtained the local fundamental systems of solutions
at $x=0,1$ and $\infty$ by using the twisted cycles.
Then, the connection relation among the local fundamental systems of solutions
can be given by linear relations among twisted cycles.

We see that there are 208 connected components in $\mathbb{R}^4\setminus S_{\mathbb{R}}$,
where $S_{\mathbb{R}}=S\cap\mathbb{R}^4$.
Each connected component can be regarded as a twisted cycle in the following way.
Since such a component $\Delta$ is connected in $\mathbb{R}^4\setminus S_{\mathbb{R}}$,
each linear function defining the integrand $u$ takes a positive or negative value on $\Delta$.
If it takes a negative value, we multiply it by $(-1)$ so that the result takes a positive value.
After this operation, we get another function $\tilde u$, which is a scalar multiple of $u$.
We define the argument of each linear function to be $0$,
which fixes the branch of $\tilde u$ on $\Delta$.
Thus $\Delta$ becomes a twisted cycle.
We call this procedure the standard loading (\cite{mimachi}).
For example, we consider the chain
$\Delta_{01a}=\{t_2<0<t_1<1<t_3<t_4\}$.
We are assuming $0<x<1$.
On this chain, $t_1,t_1-t_2,t_3,t_3-1,t_4,t_4-x$ take positive values, while
$t_1-1,t_2,t_2-1,t_2-t_3,t_3-t_4$ take negative ones.
Then we define
$$
 \begin{aligned}
  \tilde u&=  {t_1}^{\nu_1}(1-t_1)^{\nu_2-1}(t_1-t_2)^{\sigma_1-1}
  (-t_2)^{\nu_3}(1-t_2)^{\nu_4}(t_3-t_2)^{\sigma_2}
  {t_3}^{-\nu_3-\sigma_2-1}\\
  &\ \times
  (t_3-1)^{-\nu_4-\sigma_2-1}(t_4-t_3)^{-\sigma_1+\sigma_2}{t_4}^{-\nu_1-\sigma_2-1}
  (t_4-x)^{-\nu_2+\sigma_2},
 \end{aligned}
$$
where $\arg t_1=\arg(1-t_1)=\cdots=\arg(t_4-x)=0$.
We understand that the twisted cycle $\Delta_{01a}$ is defined by
the topological chain $\Delta_{01a}$
together with the above branch of $\tilde u$ on the chain.
We denote by the same letter $\Delta_{01a}$ the integral of $\tilde u$ over the twisted cycle
$\Delta_{01a}$.
In this way, we obtain 208 twisted cycles, and hence the same number of integrals.

Now we shall obtain linear relations among 208 integrals.
We set
$$
 \begin{aligned}
  e_1&=e^{\pi i\nu_1},\
  e_2=e^{\pi i(\nu_2-1)},\
  e_3=e^{\pi i(\sigma_1-1)},\
  e_4=e^{\pi i \nu_3},\
  e_5=e^{\pi i \nu_4},\
  e_6=e^{\pi i \sigma_2},\\
  e_7&=e^{\pi i(-\nu_3-\sigma_2-1)},\
  e_8=e^{\pi i (-\nu_4-\sigma_2-1)},\
  e_9=e^{\pi i (-\sigma_1+\sigma_2)},\
  e_{10}=e^{\pi i (-\nu_1-\sigma_2-1)},\\
  e_{11}&=e^{\pi i (-\nu_2+\sigma_2)}.
 \end{aligned}
$$
Then we have the relations
$$
 e_7=-\frac{1}{e_4e_6},\
 e_8=-\frac{1}{e_5e_6},\
 e_9=-\frac{e_6}{e_3},\
 e_{10}=-\frac{1}{e_1e_6},\
 e_{11}=-\frac{e_6}{e_2}.
$$
However, in order to avoid the difficulty coming from the resonance,
for a moment we assume that $e_1,e_2,\dots,e_{11}$ are algebraically independent.
On this assumption, we can obtain linear relations among the integrals by using
Cauchy's theorem.
For example, we consider $\Delta_{03a},\Delta_{03b},\Delta_{03c},\Delta_{03d}$
in 
(\ref{rank6sol56}).
Take $(t_2,t_3,t_4)\in\mathbb{R}^3$ satisfying
$t_2>1,0<t_3<t_4<x$, and fix it.
The integrand $\tilde u$ for $\Delta_{03a}$ with $(t_2,t_3,t_4)$ fixed is holomorphic
in the complex $t_1$ upper half plane and lower half plane.
Then by Cauchy's theorem, the integral of $\tilde u$ over simple loops in the upper and lower
half planes become $0$.
The loops can be deformed to unions of the intervals $(-\infty,0),(0,1),(1,t_2)$ and $(t_2,+\infty)$.
Each interval yields the chains $\Delta_{03a},\Delta_{03b},\Delta_{03c},\Delta_{03d}$
by varying $(t_2,t_3,t_4)$ satisfying the above inequalities.
Thus we get the following two relations
$$
 \begin{aligned}
  &\Delta_{03a}+e_1\Delta_{03b}+e_1e_2\Delta_{03c}+e_1e_2e_3\Delta_{03d}=0\;,
  \\  &
  \Delta_{03a}+\frac{1}{e_1}\Delta_{03b}
  +\frac{1}{e_1e_2}\Delta_{03c}+\frac{1}{e_1e_2e_3}\Delta_{03d}=0\;.
 \end{aligned}
$$
In this way, we obtain many relations among 208 integrals.
By the Gauss elimination, we can solve these relations to get the connection relations.

\bigskip\noindent
{\bf Connection Theorem }
{\it
We assume $0<x<1$.
Then the local solutions at $x=1$ are expressed by the basis
$\Delta_{01a},\Delta_{01b},\Delta_{02a},\Delta_{02b},\Delta_{03a},\Delta_{03c}$
of local solutions at $x=0$ 
as follows:
 \begin{align}
  \Delta_{11a}&=\frac{e_1e_2^2(e_1^2-1)(e_{3456}^2-1)}{e_{56}(e_{12}^2-1)(e_{1234}^2-1)}
  \Delta_{01a}
  +\frac{e_1e_2^2e_4(e_3^2-1)(e_1^2-e_{56}^2)}{e_{56}(e_{12}^2-1)(e_{1234}^2-1)}
  \Delta_{01b}\nonumber \\
  &\ 
  -\frac{(e_3^2-1)(e_2^2-e_6^2)(e_{123456}^2-1)}
  {e_{256}(e_{12}^2-1)(e_{34}^2-1)(e_3^2-e_6^2)}
  \Delta_{02a}
  +\frac{e_{134}(e_3^2-1)(e_2^2-e_6^2)(e_{256}^2-1)}
  {e_{26}(e_{12}^2-1)(e_{34}^2-1)(e_3^2-e_6^2)}
  \Delta_{02b}\nonumber \\
  &\ 
  +\frac{e_2e_3^3e_4(e_1^2-1)(e_2^2-e_6^2)(e_{456}^2-1)}
  {e_5e_6^2(e_{23}^2-1)(e_{34}^2-1)(e_{1234}^2-1)}
  \Delta_{03a}
  +\frac{e_{134}(e_3^2-1)(e_2^2-e_6^2)(e_{23456}^2-1)}
  {e_5e_6^2(e_{23}^2-1)(e_{34}^2-1)(e_{1234}^2-1)}
  \Delta_{03c},\nonumber \\
  \Delta_{11b}&=
  -\frac{e_1e_2^2e_4(e_1^2-1)(e_3^2-e_6^2)}{e_6(e_{12}^2-1)(e_{1234}^2-1)}
  \Delta_{01a}
   -
  \frac{e_1e_2^2(e_{14}^2-1)(e_3^2-e_6^2)}{e_6(e_{12}^2-1)(e_{1234}^2-1)}
  \Delta_{01b}\nonumber \\
  &\
  +\frac{e_4(e_{123}^2-1)(e_2^2-e_6^2)}{e_{26}(e_{12}^2-1)(e_{34}^2-1)}
  \Delta_{02a}
  -
  \frac{e_{13}(e_2^2-e_4^2)(e_2^2-e_6^2)}{e_{26}(e_{12}^2-1)(e_{34}^2-1)}
  \Delta_{02b}\nonumber \\
  &\
  -\frac{e_{13}e_4^2(e_2^2-e_6^2)(e_3^2-e_6^2)}{e_6^2(e_{34}^2-1)(e_{1234}^2-1)}
  \Delta_{03c},\nonumber \\
  \Delta_{11c}&=
  -\frac{e_{12}(e_3^2-e_6^2)}{e_{36}(e_{12}^2-1)}
  \Delta_{01b}
  +
  \frac{e_1(e_2^2-e_6^2)}{e_6(e_{12}^2-1)}
  \Delta_{02b},\nonumber \\
  \Delta_{11d}&=
  \frac{e_{12}(e_{34}^2-1)}{e_{1234}^2-1}
  \Delta_{01a}
  +
  \frac{e_{124}(e_3^2-1)}{e_{1234}^2-1}
  \Delta_{01b}\nonumber \\
  &\
  +
  \frac{e_{34}(e_{123}^2-1)(e_2^2-e_6^2)}{e_6(e_{23}^2-1)(e_{1234}^2-1)}
  \Delta_{03a}
  +
  \frac{e_{1234}(e_3^2-1)(e_2^2-e_6^2)}{e_6(e_{23}^2-1)(e_{1234}^2-1)}
  \Delta_{03c},
 \label{lin_rel_1}
 \end{align}
 \begin{align}
  \Delta_{12b}&=
  -\frac{e_1(e_2^2-1)(e_{2345}^2-1)(e_{123}^2-e_6^2)}
  {e_{35}(e_{12}^2-1)(e_{1234}^2-1)(e_2^2-e_6^2)}
  \Delta_{01a}
  -
  \frac{e_{14}(e_2^2-1)(e_{235}^2-1)}{e_{35}(e_{12}^2-1)(e_{1234}^2-1)(e_2^2-e_6^2)}
  \Delta_{01b}\nonumber \\
  &\
  -\frac{(e_{123}^2-1)(e_{2345}^2-1)}{e_{235}(e_{12}^2-1)(e_{34}^2-1)}
  \Delta_{02a}
  +
  \frac{e_{14}(e_2^2-1)(e_{235}^2-1)}{e_{25}(e_{12}^2-1)(e_{34}^2-1)}
  \Delta_{02b}\nonumber \\
  &\
  +\frac{e_2e_3^2e_4(e_{123}^2-1)(e_5^2-1)(e_{46}^2-1)}
  {e_{56}(e_{23}^2-1)(e_{34}^2-1)(e_{1234}^2-1)}
  \Delta_{03a}
  -
  \frac{e_1e_3^2e_4(e_2^2-1)(e_{235}^2-1)(e_{46}^2-1)}
  {e_{56}(e_{23}^2-1)(e_{34}^2-1)(e_{1234}^2-1)}
  \Delta_{03c},\nonumber \\
  \Delta_{12d}&=
  \frac{e_2(e_1^2-1)(e_{123345}^2-1)}{e_{35}(e_{12}^2-1)(e_{1234}^2-1)}
  \Delta_{01a}
  +
  \frac{e_1^2e_{24}(e_3^2-1)(e_{235}^2-1)}{e_{35}(e_{12}^2-1)(e_{1234}^2-1)}
  \Delta_{01b}\nonumber \\
  &\
  -\frac{(e_3^2-1)(e_{123345}^2-1)(e_{16}^2-1)}{e_{135}(e_{12}^2-1)(e_{34}^2-1)(e_3^2-e_6^2)}
  \Delta_{02a}
  +
  \frac{e_4(e_3^2-1)(e_{235}^2-1)(e_{16}^2-1)}{e_5(e_{12}^2-1)(e_{34}^2-1)(e_3^2-e_6^2)}
  \Delta_{02b}\nonumber \\
  &\
  +\frac{e_{23}^2e_4(e_1^2-1)(e_5^2-1)(e_{1346}^2-1)}
  {e_{156}(e_{23}^2-1)(e_{34}^2-1)(e_{1234}^2-1)}
  \Delta_{03a}
  +
  \frac{e_{24}(e_3^2-1)(e_{235}^2-1)(e_{1346}^2-1)}
  {e_{56}(e_{23}^2-1)(e_{34}^2-1)(e_{1234}^2-1)}
  \Delta_{03c},
 \label{lin_rel_2}
 \end{align}

 \begin{align}
  \Delta_{13}&=
  -\frac{e_2^2(e_1^2-1)(e_{1234}^2-e_6^2)(e_{3456}^2-1)}
  {e_{3456}(e_{12}^2-1)(e_{1234}^2-1)(e_2^2-e_6^2)}
  \Delta_{01a}
  +
  \frac{e_2^2X}{e_{356}(e_{12}^2-1)(e_{1234}^2-1)(e_2^2-e_6^2)}
  \Delta_{01b} \nonumber \\
  &\
  +\frac{(e_{123}^2-1)(e_{34}^2-e_6^2)(e_{123456}^2-1)}
  {e_{123456}(e_{12}^2-1)(e_{34}^2-1)(e_3^2-e_6^2)}
  \Delta_{02a}
  +
  \frac{Y}{e_{256}(e_{12}^2-1)(e_{34}^2-1)(e_3^2-e_6^2)}
  \Delta_{02b}\nonumber \\
  &\
  +\frac{e_2e_{34}^2(e_1^2-1)(e_{123}^2-1)(e_5^2-1)}
  {e_{15}(e_{23}^2-1)(e_{34}^2-1)(e_{1234}^2-1)}
  \Delta_{03a}
  +
  \frac{Z}{e_5e_6^2(e_{23}^2-1)(e_{34}^2-1)(e_{1234}^2-1)}
  \Delta_{03c},
 \label{lin_rel_3}
 \end{align}
where
$$
 \begin{aligned}
  X&=
  -e_{123}^2+e_{11234}^2+e_6^2-e_{16}^2+e_{136}^2-e_{1346}^2+e_{12356}^2
  +e_{123456}^2+e_{1233456}^2\\
  &\
  -e_{1123456}^2-e_{3566}^2+e_{134566}^2,\\
  Y&=
  e_{23}^2-e_{34}^2+e_6^2-e_{26}^2-e_{1236}^2+e_{12346}^2-e_{2356}^2
  +e_{23456}^2+e_{1233456}^2\\
  &\
  -e_{12233456}^2+e_{122356}^2-e_{1234566}^2,\\
  Z&=
  -1+e_{23}^2+e_{346}^2-e_{2346}^2+e_{12346}^2-e_{123346}^2+e_{23456}^2
  -e_{233456}^2 +e_{1233456}^2\\
  &\
 -e_{12233456}^2-e_{123344566}^2
  +e_{12233344566}^2.
 \end{aligned}
$$
}

\bigskip\noindent
{\bf Corollary}
{\it
We have a linear dependence relation
$$
 \Delta_{11a}+c_2\Delta_{11b}+c_3\Delta_{11c}+c_4\Delta_{11d}=0
$$
among the regularizable cycles of exponent $0,1,2$ at $x=1$,
where
$$
 \begin{aligned}
  c_2&=\frac{(e_3^2-1)(e_{123456}^2-1)}{e_{45}(e_{123}^2-1)(e_3^2-e_6^2)},\\
  c_3&=-\frac{e_{23}(e_3^2-1)(e_{134}^2-1)(e_{456}^2-1)}
  {e_{45}(e_{123}^2-1)(e_{34}^2-1)(e_3^2-e_6^2)},\\
  c_4&=-\frac{e_2e_3^2(e_1^2-1)(e_{456}^2-1)}{e_{56}(e_{123}^2-1)(e_{34}^2-1)}.
 \end{aligned}
$$
If $e_1,e_2,\dots,e_6$ are algebraically independent,
there is no other linear relation among $\Delta_{11a},\Delta_{11b},\Delta_{11c},\Delta_{11d}$.
}

\bigskip
{\it Proof.}
We expressed above  
the integrals (twisted cycles) $\Delta_{11a},\allowbreak\Delta_{11b},\Delta_{11c},\Delta_{11d}$
as linear combinations of 
$\Delta_{01a},\Delta_{01b},\Delta_{02a},\Delta_{02b},\Delta_{03a},\Delta_{03c}$.
Denote the $6\times4$ matrix of coefficients of the linear combinations by $A$.
Then we have
$$
 A\begin{pmatrix}1\\ c_2\\ c_3\\ c_4\end{pmatrix}=\vec 0\;,
$$
and that the rank of $A$ is $3$ if $e_1,e_2,\dots,e_6$ are algebraically independent.
$\square$

\subsection{Connection matrix, Monodromy and invariant Hermitian form}

Thanks to Corollary, we can take three of $\Delta_{11a},\Delta_{11b},\Delta_{11c},\Delta_{11d}$
as a basis of the space of holomorphic solutions at $x=1$.

\noindent Thus, for example, we have a basis
$\Delta_{11b},\Delta_{11c},\Delta_{11d},\Delta_{12b},\allowbreak\Delta_{12d},\Delta_{13}$
of local solutions at $x=1$.
The (\ref{lin_rel_1}),(\ref{lin_rel_2}) and (\ref{lin_rel_3})  give the connection relation between the local bases at $x=0$ and at $x=1$,
and hence we can derive the monodromy representation of the differential equation
with the Riemann scheme (\ref{rank6RS}).
We take $\mathcal{Y}_0(x)=(\Delta_{01a},\Delta_{01b},\Delta_{02a},\Delta_{02b},\Delta_{03a},\Delta_{03c})$ as
the basis of  local solutions at $x=0$,
and $\mathcal{Y}_1(x)=(\Delta_{11b},\Delta_{11c},\Delta_{11d},\Delta_{12b},\Delta_{12d},\Delta_{13})$
as one at $x=1$.
The connection matrix $C$ defined by
\begin{equation}
\label{connection}
 \mathcal{Y}_1(x)=\mathcal{Y}_0(x)C
\end{equation}
is read by (\ref{lin_rel_1}),(\ref{lin_rel_2}) and (\ref{lin_rel_3}).
Let $\gamma_0$ and $\gamma_1$ be loops with base point $x=1/2$ which encircle $x=0$ and $x=1$,
respectively, once in the positive direction.
We denote by ${\gamma_j}_*$ the analytic continuation along $\gamma_j$ $(j=0,1)$.
Then obviously, we have
$$
 {\gamma_0}_*\mathcal{Y}_0(x)=
 \mathcal{Y}_0(x)E_0\;,\
 {\gamma_1}_*\mathcal{Y}_1(x)=
 \mathcal{Y}_0(x)E_1\;,
$$
where
\begin{equation}
\label{E0E1}
 \begin{aligned}
  E_0&=
  {\rm diag}\left[1,1,\frac{1}{{e_{12}}^2},\frac{1}{{e_{12}}^2},\frac{1}{{e_{1234}}^2},
  \frac{1}{{e_{1234}}^2}\right],\\
  E_1&=
  {\rm diag}\left[1,1,1,\frac{{e_6}^2}{{e_{235}}^2},\frac{{e_6}^2}{{e_{235}}^2},
  {e_6}^4\right].
 \end{aligned}
\end{equation}
Thus we obtain the following result.

\bigskip\noindent
{\bf Monodromy Theorem }
{\it
Let $M_0,M_1$ be the monodromy matrices for the loops $\gamma_0,\gamma_1$,
respectively, with respect to the fundamental set of solutions $\mathcal{Y}_0(x)$
of the differential equation with Riemann scheme (\ref{rank6RS}).
Then we have
$$
 M_0=E_0\;,\
 M_1=CE_1C^{-1}\;,
$$
where $C$ is defined by (\ref{connection}) and $E_0,E_1$ are given by (\ref{E0E1}).
}

\bigskip
The explicit
form of the monodromy matrix $M_1$ can be computed by using the (\ref{lin_rel_1}),
(\ref{lin_rel_2}), (\ref{lin_rel_3}) and (\ref{E0E1}).
By using the explicit form of $M_0,M_1$,
we can compute the monodromy invariant Hermitian form which 
is given by the Hermitian matrix $H$ of size $6$ satisfying
$\bar M_0H M_0^T=H,\ \bar M_1 H M_1^T=H$.
Then we find the following result.

\bigskip\noindent
{\bf Unicity Theorem }
{\it
Assume that the parameters $\nu_1,\nu_2,\nu_3,\nu_4,\sigma_1,\sigma_2$ in the Riemann scheme (\ref{rank6RS})
are real numbers.
Then the monodromy invariant Hermitian form exists
uniquely up to multiplication by $\mathbb{R}^{\times}$.
The explicit form of the Hermitian matrix $H$ is given by
\begin{equation}
\label{rank6Hermite}
H=\begin{pmatrix}  
h_{11}& h_{12}& 0&0 &0&0   \\ 
h_{12}   &h_{22} &  0   &0 & 0&0   \\
0  &0 &  h_{33}   &h_{34} & 0&0   \\
0   &0 &  h_{34} & h_{44}& 0&0\\
  0   &0 &  0 & 0& h_{55}& h_{56} \\
  0   &0 &  0 & 0& h_{56}&h_{66}
\end{pmatrix},
\end{equation}
with
\begin{equation}
\begin{aligned}
\frac{h_{11}}{h_{56}}&=
\frac{e_{12}(e_{23}^2-1)^2(e_{34}^2-1)(e_{3456}^2-1)\xi_{11}}
{e_{34}^4(e_{12}^2-1)(e_3^2-1)(e_{123}^2-1)(e_5^2-1)(e_{235}^2-1)(e_2^2-e_6^2)^2}\;,\\
\frac{h_{12}}{h_{56}}&=
\frac{e_{12}(e_{23}^2-1)^2(e_{34}^2-1)(e_{1234}^2-e_6^2)(e_{3456}^2-1)}
{e_3^4e_4^3(e_{12}^2-1)(e_{123}^2-1)(e_5^2-1)(e_2^2-e_6^2)^2}\;,\\
\frac{h_{22}}{h_{56}}&=
\frac{e_{12}(e_{23}^2-1)^2(e_{34}^2-1)\xi_{22}}
{e_3^4e_4^2(e_1^2-1)(e_{12}^2-1)(e_{123}^2-1)(e_5^2-1)(e_2^2-e_6^2)^2}\;,\\
\frac{h_{33}}{h_{56}}&=
\frac{(e_{23}^2-1)^2(e_{1234}^2-1)(e_{123456}^2-1)\xi_{33}}
{e_1e_2^3e_{34}^4(e_1^2-1)(e_2^2-1)(e_{12}^2-1)(e_5^2-1)(e_{235}^2-1)(e_3^2-e_6^2)^2}\;,\\
\frac{h_{34}}{h_{56}}&=
-\frac{(e_{23}^2-1)^2(e_{1234}^2-1)(e_{34}^2-e_6^2)(e_{123456}^2-1)}
{e_{234}^3(e_1^2-1)(e_{12}^2-1)(e_5^2-1)(e_3^2-e_6^2)^2}\;,\\
\frac{h_{44}}{h_{56}}&=
\frac{e_1(e_{23}^2-1)^2(e_{1234}^2-1)\xi_{44}}
{e_2^3e_{34}^2(e_1^2-1)(e_{12}^2-1)(e_{123}^2-1)(e_5^2-1)(e_3^2-e_6^2)^2}\;,\\
\frac{h_{55}}{h_{56}}&=
\frac{e_2\xi_{55}}
{e_1e_{46}^2(e_2^2-1)(e_3^2-1)(e_{235}^2-1)}\;,\\
\frac{h_{66}}{h_{56}}&=
\frac{e_1\xi_{66}}
{e_2e_{346}^2(e_1^2-1)(e_{123}^2-1)(e_5^2-1)}\;,\\
\end{aligned}
\label{Hrank6}
\end{equation}
where
$$
\begin{aligned}
\xi_{11}
&=e_{23}^2-e_{234}^2+e_{1234}^2-e_{124}^2e_3^4-e_{145}^2e_{23}^4
+e_{15}^2e_{24}^4e_3^6-e_6^2+e_{346}^2+e_{23456}^2-e_{2456}^2e_3^4\\
&\quad
+e_{12456}^2e_3^4-e_{1256}^2e_{34}^4\;,\\
\xi_{22}&=e_{123}^2-e_1^4e_{234}^2-e_6^2+e_{16}^2-e_{136}^2+e_{1346}^2
-e_{12356}^2+e_{123456}^2-e_{12456}^2e_3^4+e_{13}^4e_{2456}^2\\
&\quad
+e_{35}^2e_6^4-e_{1245}^2e_6^4\;,\\
\xi_{33}
&=e_{23}^2+e_{34}^2-e_{234}^2-e_{124}^2e_3^4-e_{245}^2e_3^4+e_{15}^2e_{24}^4e_3^6-e_6^2+e_{12346}^2+e_{23456}^2+e_{12456}^2e_3^4\\
&\quad-e_{1456}^2e_{23}^4-e_{1256}^2e_{34}^4\;,\\
\xi_{44}
&=-e_{23}^2+e_{34}^2-e_6^2+e_{26}^2+e_{1236}^2-e_{12346}^2
+e_{2356}^2-e_{23456}^2-e_{12456}^2e_3^4+e_{1456}^2e_{23}^4\\
&\quad
-e_{135}^2e_{26}^4+e_{12345}^2e_6^4\;,\\
\xi_{55}
&=-1+e_{23}^2+e_{46}^2-e_{346}^2+e_{1346}^2-e_{12346}^2
+e_{3456}^2-e_{23456}^2+e_{123456}^2-e_{12456}^2e_3^4\\
&\quad
-e_{135}^2e_{46}^4+e_{125}^2e_{346}^4\;,\\
\xi_{66}
&=-1+e_{23}^2+e_{346}^2-e_{2346}^2+e_{12346}^2-e_{1246}^2e_3^4
+e_{23456}^2-e_{2456}^2e_3^4+e_{12456}^2e_3^4-e_{1456}^2e_{23}^4\\
&\quad-e_{125}^2e_{346}^4+e_{15}^2e_{246}^4e_3^6\;.
\end{aligned}
$$
}

\subsection{$\cW_3$ correlation function and structure constants}

We use the above results to compute the $\cW_3$ correlation function.
We showed that the integral (\ref{rank6Integral}) is a solution of the sixth-order ODE found in \cite{BCES17} and therefore it provides, via the relation (\ref{rank6coeff}),  an integral representation of the conformal block $\cB_{M}^{(1,s)}$. 
In particular, having found the basis 
$\mathcal{Y}_0(x)$ of local solutions at $x=0$, see (\ref{rank6sol12}), (\ref{rank6sol34}) and (\ref{rank6sol56}), we can do the following associations:
\begin{align}
\Delta_{01a},\Delta_{01b} &\to \text{Fusion \ref{fusion1}}\to \cB_{M_1}^{(1,s)} \nonumber \\
\Delta_{02a},\Delta_{02b} &\to \text{Fusion \ref{fusion2}} \to \cB_{M_2}^{(1,s)} \nonumber \\
\Delta_{03a},\Delta_{03c} &\to \text{Fusion \ref{fusion3}} \to \cB_{M_3}^{(1,s)} 
\end{align}
We recall that the conformal block $\cB_{M_i}^{(1,s)}$ has an additional parameter,c$\lambda^{1}_{\vec{\alpha}_L,\vec{\alpha}_{M_i}}(-b\vec{\omega}_1+s\vec{\omega}_2)$, see (\ref{param}), whose value is fixed  by specifying the linear combination between the two solutions $(\Delta_{0ia},\Delta_{0ib})$, $i=1,2,3$. 
\noindent The monodromy invariant  form $H$  provides explicitly the factorization (\ref{corr_decomp}) of the correlation function:
\begin{equation}
\label{factrank6}
 \left\langle \Phi_{\vec{\alpha}_{L}}(\infty)\Phi_{-b\vec{\omega}_1+s\vec{\omega}_2}(1)
 \Phi_{-b \vec{\omega}_1}(x)\Phi_{\vec{\alpha}_{R}}(0)\right\rangle
  = \mathcal{Y}^{*}_0(x) H \mathcal{Y}_0^T(x)\; ,
\end{equation}
where $H$ is given by (\ref{rank6Hermite}). As we stated in the unicity theorem, the coefficients $h_{ij}$ entering the $H$ matrix are uniquely determined as a function of one parameter, say $h_{56}$ (this can be related to the freedom of a global normalization of the correlation function). It is important to remark that the correlation function in (\ref{factrank6}) has been studied in detail in \cite{fl08} where a Coulomb gas approach was used to provide an eight dimensional integral representation. In this respect our solution provides the factorization of this eight dimensional integral in terms of holomorphic and anti-holomorphic four dimensional integrals. We stress that this factorization is highly non-trivial as can be seen from the fact that the absolute square of the integrand  (\ref{rank6Integral}) do not coincide with the integrand appearing in the  solutions of \cite{fl08}.
However, we could prove the equivalence between our results and the one in \cite{fl08}  by comparing the values of the structure constants.
Our results allow us to compute indeed new shift relations for the structure constants  $C(\vec{\alpha}_{L}, -b\vec{\omega}_1+s\vec{\omega}_2, \alpha_{M_i})$. 
In order to do that, we need to evaluate the initial values  
$\mathcal{N}_{X}$, $X=01a,01b,02a,02b,03a,03c$,
of the Taylor expansions at  $x=0$ in $\mathcal{Y}_0(x)$, i.e.  
$$
 \begin{aligned}
  \mathcal{Y}_0(x)
  &=
  \bigl(\mathcal{N}_{01a}+O(x),\mathcal{N}_{01b}+O(x),
  x^{-\nu_{12}}(\mathcal{N}_{02a}+O(x)),x^{-\nu_{12}}(\mathcal{N}_{02b}+O(x)),\\
  &\qquad
  x^{-\nu_{123}-\sigma_1}(\mathcal{N}_{03a}+O(x)),
  x^{-\nu_{123}-\sigma_1}(\mathcal{N}_{03c}+O(x))\bigr).
 \end{aligned}
$$
Four of these constants are more conveniently given by the 3-dimensional integrals:
\begin{equation}
\label{nX}
\mathcal{N}_{X}=\int_{0}^{1} d t_1 \int_{0}^{1} d t_2 \int_{0}^{1} d t_3  \; \phi_{X}(t_1,t_2,t_3)\;,\quad  X=01a,01b,02a,02b
\end{equation}
with:
\begin{eqnarray}
 \phi_{01a}(t_1,t_2,t_3)&=& 
 \frac{\Gamma(\nu_{12} - \sigma_2 + \sigma_1)\Gamma(1-\sigma_1+\sigma_2)}
 {\Gamma(1+\nu_{12})}
 \left[t_1^{\nu_1} (1 - t_1)^{\nu_2 - 1} (1 - t_2)^
 {\nu_3} \;t_2^{-1 - \nu_{34} - \sigma_{12}} \times \right. \nonumber \\  
 & & \left. \times (t_2 t_1 - t_2 + 1)^{\sigma_1 - 1} 
 (t_2 - t_2 t_3 + t_3)^{\sigma_2}
 \;  t_3^{\nu_{1234} + \sigma_1} (1 - t_3)^{-\nu_4 - \sigma_2 - 1}\right]\;,\\
 \phi_{01b}(t_1,t_2,t_3)&=& \frac{\Gamma(\nu_{12} - \sigma_2 + \sigma_1)
 \Gamma(1-\sigma_1+\sigma_2)}
 {\Gamma(1+\nu_{12})}\left[t_1^{\nu_{13}+\sigma_1} (1 - t_1)^{\nu_2 - 1} t_2^{\nu_3}(1 - t_2)^
 {\sigma_1-1} \; \times \right. \nonumber \\  & & \left. \times (1 - t_1 t_2)^{\nu_4} 
 (1 - t_1 t_2 t_3)^{\sigma_2}
 \;  t_3^{\nu_{1234} + \sigma_1} (1 - t_3)^{-\nu_4 - \sigma_2 - 1}\right]\;,\\
\phi_{02a}(t_1,t_2,t_3)&=& \frac{\Gamma(-\nu_1  - \sigma_2 )\Gamma(1-\nu_2+\sigma_2)}
{\Gamma(1-\nu_{12})}\left[t_1^{-1-\nu_{1234}-\sigma_{12}} (1 - t_1)^{\nu_{13}+\sigma_1} (1 - t_2)^
 {\sigma_1-1} \;t_2^{\nu_3} \times \right. \nonumber \\  & & \left. \times (t_1+ t_2 - t_1 t_2 )^{\nu_4} 
 (t_1 +t_2 t_3- t_1 t_2 t_3)^{\sigma_2}
 \;  t_3^{\nu_{34}+ \sigma_1} (1 - t_3)^{-\nu_4 - \sigma_2 - 1}\right]\;,\\    
 \phi_{02b}(t_1,t_2,t_3)&=& \frac{\Gamma(-\nu_1  - \sigma_2 )\Gamma(1-\nu_2+\sigma_2)}
 {\Gamma(1-\nu_{12})}\left[t_1^{\nu_{13}+\sigma_1} (1 - t_1)^{\nu_2-1} t_2^{\nu_3} (1 - t_2)^
 {\sigma_1-1}  \times \right. \nonumber \\  & & \left. \times (1 - t_1 t_2 )^{\nu_4} 
 (1- t_1 t_2 t_3)^{\sigma_2}
 \;  t_3^{\nu_{34}+ \sigma_1} (1 - t_3)^{-\nu_4 - \sigma_2 - 1}\right]\;.
\label{N1234}
    \end{eqnarray}
The remaining constants $\mathcal{N}_{03a}$, $\mathcal{N}_{03c}$ can be given in terms of
hypergeometric function ${}_3 F_{2}[p_1,p_2,p_3;q_1,q_2;x]$  evaluated at $x=1$:
\begin{eqnarray}
\mathcal{N}_{03a} &=&  \frac{\Gamma (-\nu_{13}-\sigma_{12}) \Gamma (1-\nu_2+\sigma_2)}{\Gamma (1-\nu_{123}-\sigma_1)}\frac{\Gamma (-\nu_3-\sigma_2) \Gamma (\sigma_2-\sigma_1 +1)}{\Gamma (-\nu_3-\sigma_1 +1)}\times \nonumber\\
&&\times \;{}_3 F_{2}[1-\sigma_1,1+\nu_3,1+\nu_1;1-\nu_3-\sigma_{12},2-\nu_2-\sigma_1;1] \;,\nonumber \\
\mathcal{N}_{03c} &=&  \frac{\Gamma (-\nu_2+\sigma_2+1) \Gamma (-\nu_{13}-\sigma_{12})}{\Gamma (-\nu_{123}-\sigma_1 +1)}\frac{\Gamma (-\nu_3-\sigma_2) \Gamma (\sigma_2-\sigma_1 +1)}{\Gamma (-\nu_3-\sigma_1 +1)}\times \nonumber \\
&&\times \;{}_3 F_{2}[-\nu_4, -\nu_{34}-\sigma_{12},-\nu_{1234}-\sigma_{12};-\nu_{34}-\sigma_2,-\nu_{134}-\sigma_{12};1] \;.
\label{N56}
\end{eqnarray}

\noindent We can now give our final result. Defining:
\begin{equation}
\tilde{h}_{ij} =h_{ij}\mathcal{N}_i \mathcal{N}_j\;,
\end{equation}
the ratio of structure constants that are associated to the different fusion channels take the following form: 
\begin{eqnarray}
\frac{C(2Q\vec{\rho}-\vec{\alpha}_{L}, -b\vec{\omega}_1+s\vec{\omega}_2, \vec{\alpha}_{M_1}+ b \vec{e}_1)C(\vec{\alpha}_{R}, -b\vec{\omega}_1, 2Q\vec{\rho}-\vec{\alpha}_{M_1}- b \vec{e}_1)}{C(2Q\vec{\rho}-\vec{\alpha}_{L}, -b\vec{\omega}_1+s\vec{\omega}_2, \vec{\alpha}_{M_1})C(\vec{\alpha}_{R}, -b\vec{\omega}_1, 2Q\vec{\rho}-\vec{\alpha}_{M_1})}&=&\frac{\tilde{h}_{33}+2\tilde{h}_{34}+\tilde{h}_{44}}{\tilde{h}_{11}+2\tilde{h}_{12}+\tilde{h}_{22}}\;, \nonumber\\
\frac{C(2Q\vec{\rho}-\vec{\alpha}_{L}, -b\vec{\omega}_1+s\vec{\omega}_2, \vec{\alpha}_{M_1}+ b \vec{\rho})C(\vec{\alpha}_{R}, -b\vec{\omega}_1, 2Q\vec{\rho}-\vec{\alpha}_{M_1}- b \vec{\rho})}{C(2Q\vec{\rho}-\vec{\alpha}_{L}, -b\vec{\omega}_1+s\vec{\omega}_2, \vec{\alpha}_{M_1})C(\vec{\alpha}_{R}, -b\vec{\omega}_1, 2Q\vec{\rho}-\vec{\alpha}_{M_1})}&=&\frac{\tilde{h}_{55}+2\tilde{h}_{56}+\tilde{h}_{66}}{\tilde{h}_{11}+2\tilde{h}_{12}+\tilde{h}_{22}}\;.\nonumber \\
& & \label{rec_relation}
\end{eqnarray}
Even if the values of $h_{ij}$ and of $\mathcal{N}_{X}$, see (\ref{Hrank6}), (\ref{N1234}) and (\ref{N56}),  have been given in terms of the Katz parameter $(\nu_{1}, \nu_{2},\sigma_{1},\nu_{3},\nu_{4},\sigma_{2})$, we preferred, in the l.h.s of the above equation, to keep the dependence of the structure constants on the Toda parameters $(a_{R_1},a_{R_2},a_{L_1},a_{L_2},b,s)$. We recall that these two set of parameters are simply related by (\ref{rank6coeff}). 
By applying the shift relations to the constants of the type $C(\vec{\alpha}_{1}, -b\vec{\omega}_1, \vec{\alpha}_2)$ \cite{fl07c}: 
\begin{eqnarray}
\frac{C(\vec{\alpha}_{R}, -b\vec{\omega}_1, 2Q\vec{\rho}-\vec{\alpha}_{M_1}- b \vec{e}_1)}{C(\vec{\alpha}_{R}, -b\vec{\omega}_1, 2Q\vec{\rho}-\vec{\alpha}_{M_1})} &=& \sigma_2^{1 + \sigma_2} \frac{\Gamma(\nu_{12}) \Gamma(-\nu_{12}-\sigma_2)}{(
\Gamma(1 -\nu_{12}) \Gamma(1 +\nu_{12}+\sigma_2)}\;, \nonumber \\
\frac{C(\vec{\alpha}_{R}, -b\vec{\omega}_1, 2Q\vec{\rho}-\vec{\alpha}_{M_1}- b \vec{\rho})}{C(\vec{\alpha}_{R}, -b\vec{\omega}_1, 2Q\vec{\rho}-\vec{\alpha}_{M_1})} &=& \pi^2\; \frac{\Gamma(1+\sigma_2)^2}{\Gamma(-\sigma_2)^2}\frac{
\Gamma(\nu_3+\sigma_1)\Gamma(\nu_{123}+\sigma_1)}{ \Gamma(
    1 - \nu_3 - \sigma_1) \Gamma(1 -\nu_{123} - \sigma_1)}\times \nonumber \\ &&\times   \frac{\Gamma(-\nu_3-\sigma_{12})\Gamma(-\nu_{123}-\sigma_{12})}{
\Gamma(1 + \nu_3 + \sigma_{12})\Gamma(1 + \nu_{123} + \sigma_{12})}\;,
\end{eqnarray}
the (\ref{rec_relation}) can be written in the form of shifts relations of the structure constants containing one semi-degenerate field at second level. These relations take the following form: 
\begin{align}
\frac{C(2Q\vec{\rho}-\vec{\alpha}_{L}, -b\vec{\omega}_1+s\vec{\omega}_2, \vec{\alpha}_{M_1}+ b \vec{e}_1)}{C(2Q\vec{\rho}-\vec{\alpha}_{L}, -b\vec{\omega}_1+s\vec{\omega}_2, \vec{\alpha}_{M_1})}&= \sigma_2^{-1 - \sigma_2}\frac{\tilde{h}_{33}+2\tilde{h}_{34}+\tilde{h}_{44}}{\tilde{h}_{11}+2\tilde{h}_{12}+\tilde{h}_{22}}  \frac{(
\Gamma(1 -\nu_{12}) \Gamma(1 +\nu_{12}+\sigma_2)}{\Gamma(\nu_{12}) \Gamma(-\nu_{12}-\sigma_2)}\;, \nonumber \\
\frac{C(2Q\vec{\rho}-\vec{\alpha}_{L}, -b\vec{\omega}_1+s\vec{\omega}_2, \vec{\alpha}_{M_1}+ b \vec{\rho})}{C(2Q\vec{\rho}-\vec{\alpha}_{L}, -b\vec{\omega}_1+s\vec{\omega}_2, \vec{\alpha}_{M_1})}&= \pi^{-2}\frac{\tilde{h}_{55}+2\tilde{h}_{56}+\tilde{h}_{66}}{\tilde{h}_{11}+2\tilde{h}_{12}+\tilde{h}_{22}}\; \frac{\Gamma(-\sigma_2)^2}{\Gamma(\sigma_2)^2} \times \nonumber \\ 
\frac{
\Gamma(\nu_3+\sigma_1)\Gamma(\nu_{123}+\sigma_1)}{ \Gamma(
    1 - \nu_3 - \sigma_1) \Gamma(1 -\nu_{123} - \sigma_1)}  &  \times \frac{\Gamma(1 + \nu_3 + \sigma_{12})\Gamma(1 + \nu_{123} + \sigma_{12})}{\Gamma(-\nu_3-\sigma_{12})\Gamma(-\nu_{123}-\sigma_{12})
}\;. 
\end{align}
We verified that the above formulas  are in agreement with the values of the structure constants computed in \cite{fl08}.

\section{Summary and conclusions}
\label{summary}

In this paper we discussed a family of differential equations satisfied by  special $\cW_3$ conformal blocks, defined in (\ref{cbconsidered}). 
We derived  in section (\ref{cb_sem_deg}) their Riemann scheme and we made the crucial observation that these equations are related to rigid Fuchsian systems. 
These equations have a fundamental nature as they are the simplest equations that determine the matrix elements involving semi-degenerate fields at a given level. 
Being related to rigid Fuchsian systems, Katz theory can be applied. 
As a new application, we succeeded in providing the integral expression (\ref{rank6Integral}) for the function $\cB^{(1,s)}_M(x)$, see (\ref{cbconsidered}). 
This was done by solving a sixth-order rigid Fuchsian equation via the Katz theory. 
In particular, in section \ref{cycles}, we derived the cycles  corresponding to local systems of solutions, see (\ref{rank6sol12}) - (\ref{rank6soli3}).  
The monodromy group of the equation was also computed, see (\ref{lin_rel_1})-(\ref{lin_rel_3}). These results allowed to obtain the first explicit factorization (\ref{factrank6}). 
We derived new shift relations for the three-point functions with second-level semi-degenerate fields. 
We verified that our results are  in agreement with previous results obtained in \cite{fl08}.

We have seen that in rigid Fuchsian systems a differential equation is uniquely determined by the  local properties of its solutions. In CFT terms, this means that the fusion rules of (degenerate) primary operators fix also the ODE their correlation functions satisfies. This problem is also motivated by the fact that in CFT the computation of the null-vectors and the subsequent determination of the ODE is much more complicated than determining the fusions rules. A recent discussion on these issues  can be found in \cite{MuMu17}.
In this respect, a key observation is that the rigidity index $\iota \neq 2$  (\ref{iota}) for CFT equations, such as for instance the BPZ equations of order greater than two \cite{bpz84}.  Hence the Fuchsian equations appearing in CFT are not in general rigid. It will be interesting to explore more deeply the connection between CFT-Fuchs equations and whether or not similar methods are applicable in non-rigid cases. 
Finally, we believe that  Katz theory can be useful to clarify  general classification of Fuchsian equations appearing in the context of the conformal field theories with different types of chiral algebras, such as super-conformal algebras or the $\cW_N$ algebras with $N\geq 4$ \cite{fl08,furlan2015some,FuPe17}.

\appendix

\section*{Acknowledgements}
 
We thank T. Dupic, P. Gavrylenko, N. Iorgov, O. Lisovyy , Y. Matsuo and  S. Ribault  for discussions and X. Cao, B. Estienne and O. Foda for preliminary contributions and previous results on which the study of this  manuscript is based. The second author is supported by the JSPS grants-in-aid for scientific research B, No. 15H03628.

\bibliographystyle{JHEP}
\bibliography{bibnonwy.bib}

\end{document}